# COMPREHENSIVE DESIGN AND WHOLE-CAVITY SIMULATION OF A MULTI-BEAM INDUCTIVE OUTPUT TUBE USING A 3rd HARMONIC DRIVE ON THE GRID

H.P. Freund,[1,2,3] T. Bui,[1] R.L. Ives,[1] T. Habermann,[1] and M. Read[1]
[1]Calabazas Creek Research, Inc., San Mateo, CA 94404
[2]Institute for Research in Electronics and Applied Physics, University Maryland, College Park, Maryland 20742
[3]University of New Mexico, Albuquerque, New Mexico, 87131

In this paper, we discuss the design and whole-cavity simulation of a Multi-Beam Inductive Output Tube (MBIOT) that uses a 3rd harmonic component to the drive voltage on the grid. High-efficiency inductive output tubes (IOTs) are typically characterized by efficiencies up to 70 – 75%. However, the achievement of efficiencies greater than 80% would substantially reduce the operating costs of next-generation accelerators. To achieve this goal, we consider the addition of a 3rd harmonic component to the drive signal on the grid. We anticipate that the MBIOT will be used to provide the rf power to drive RF linacs. We discuss and model an 8-beam MBIOT with a 700 MHz resonant cavity using beams with a voltage of 35 kV and an average current of 7.25 A yielding a perveance of about 1.1 μP. We simulate this MBIOT using the NEMESIS simulation code which has been extended using a three-dimensional Poisson solver based upon the Petsc package from Argonne National Laboratory. The effect of the 3rd harmonic on the efficiency is greatest when the phase of the 3rd harmonic is shifted by $\pi$ radians with respect to the fundamental drive signal and with 3rd harmonic powers greater than about 50% of the fundamental drive power. For the present example, we show that efficiencies approaching 82% are possible. Designs for the MBIOT input coupler, grids and output cavity have been developed based on these simulations and will be discussed.

## I. INTRODUCTION

High-efficiency inductive output tubes (IOTs) are typically characterized by efficiencies up to 70 – 75%. However, the achievement of efficiencies greater than 80% would substantially reduce the operating costs of next-generation radio frequency linear accelerators (RF linacs). To achieve this goal, we consider the addition of a 3rd harmonic component to the drive signal on the grid of a multi-beam inductive output tube (MBIOT) [1,2]. The 3rd harmonic drive component in IOT guns has been considered to apply such a gun as the injector [3] of radio frequency linear accelerators (RF linacs).

IOTs are sometimes referred to as klystrodes® [4] and are used for high power applications such as the drivers for RF Linacs [5-8]. A schematic illustration of an IOT is shown in Fig. 1 [9]. In an IOT, a pre-bunched beam is generated by applying a radio-frequency signal to the grid of an electron gun which is accelerated to higher energies by a DC potential before injection into a resonant output cavity. The modulated beam is preconditioned to excite the resonant mode of the output cavity, after which the spent beam is directed into a collector.

Recent development of a high-power MBIOT [10] intended for the European Spallation Neutron Source [11] achieved an efficiency of 65% with a peak power of 1.2 MW at a frequency of 705 MHz. The cost of RF power for such facilities is an important driver for accelerator operations, and the production of an IOT with still higher efficiencies could provide significant cost reductions. A 5% increase in the efficiency reduces the power required for a 200-kW device by 10 kW. This is a significant operational cost savings for systems running continuously or at high duty. This is the prime motivation for studying the application of the 3rd harmonic component to the drive on the IOT gun(s).

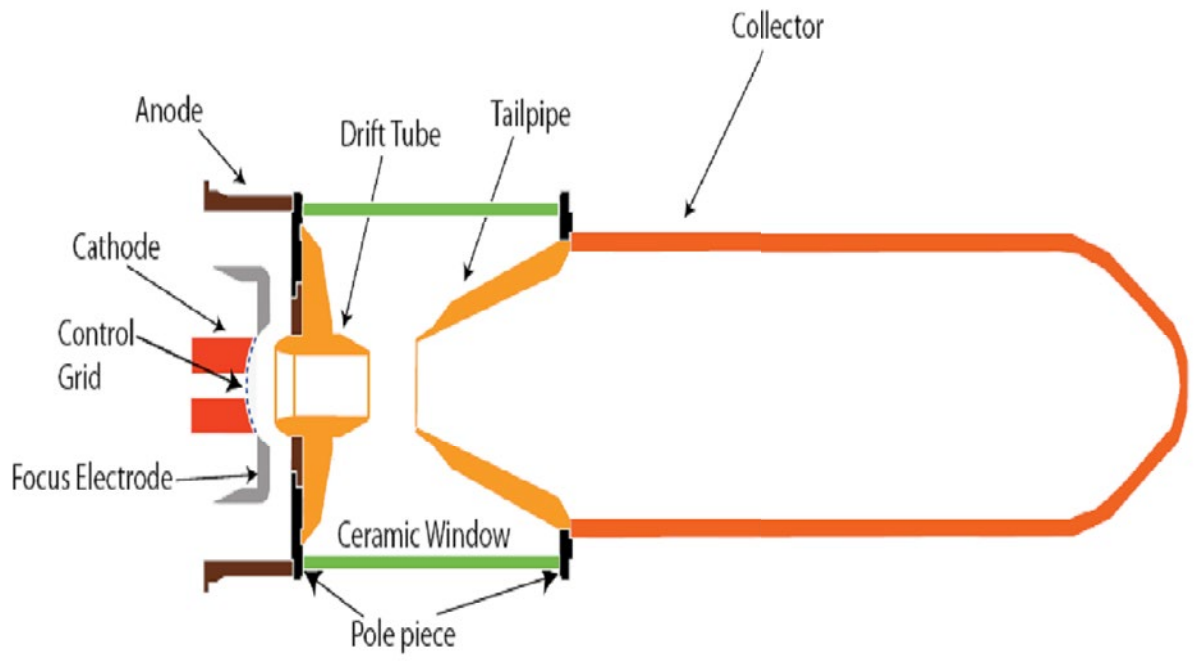

Fig. 1: Schematic illustration of an inductive output tube, where the output cavity is located between the pole pieces [9].

We consider an 8-beam MBIOT with a 700 MHz resonant cavity and using an annular beam with a voltage of 35 kV, an average current of 7.25 A yielding a perveance of about 1.1 μP. This MBIOT is simulated using the NEMESIS simulation code [1,9] which has been extended by the incorporation of a three-dimensional Poisson solver using the Petsc package available from Argonne National Laboratory [12]. NEMESIS has previously been successfully validated by comparison with the K5H90W-2 IOT developed by Communications & Power Industries LLC (CPI) [9]. It is found that the effect of the 3rd harmonic on the efficiency is greatest when the phase of the 3rd harmonic is shifted by $\pi$ radians with respect to the fundamental drive signal and with 3rd harmonic powers greater than about 50% that of the fundamental drive power. For the present example, we show that efficiencies approaching 82% are possible.

The organization of the paper is as follows. The numerical model incorporated in the NEMESIS code and whole-cavity simulations with NEMESIS are described in Section. II. The designs for the input coupler, grids and output cavity are described in Section II. A summary and discussion are given in Section IV.

## II. WHOLE-CAVITY SIMULATION

The numerical formulation in the NEMESIS code [1,9] is similar to particle-in-cell simulation codes. Integration of the dynamical equations is performed in time, so the code can implicitly treat particles that might turn around. This can be important in high-efficiency designs where particles lose a great deal of energy.

NEMESIS contains an equivalent (LRC) circuit model for the cavity voltage with a model for the circuit fields taken from Kosmahl and Branch [13] and scaled using the cavity voltage. The integration of particle trajectories uses the circuit fields obtained in this fashion as well as an analytic model for the focusing fields and two- or three-dimensional Poisson solvers for the space charge fields.

The numerical procedure is illustrated in Fig. 2. The procedure in stepping from $t \rightarrow t + \Delta t$ begins with a 4$^{th}$ order Runge-Kutta integration of the equivalent circuit equations. We typically take 100 steps per wave period to ensure accuracy. Once the circuit equations have been stepped, electrons are injected into the cavity if the time coincides with the bunch phase. After injection, the trajectories are integrated using a Boris push [14], which includes the RF and magnetostatic fields, and the Poisson solver is called to obtain the electrostatic fields. At this point, the source current is calculated by averaging the current obtained using ($\mathbf{x}_t$, $\mathbf{v}_{t+\Delta t/2}$) and ($\mathbf{x}_{t+\Delta t}$, $\mathbf{v}_{t+\Delta t/2}$). This ensures 2$^{nd}$ order accuracy for the overall procedure. Finally, we test whether any particles have left the system (from either end of the cavity or by striking the wall), and, if necessary, eject them from the simulation. This is repeated as many times as necessary to simulate any given pulse time.

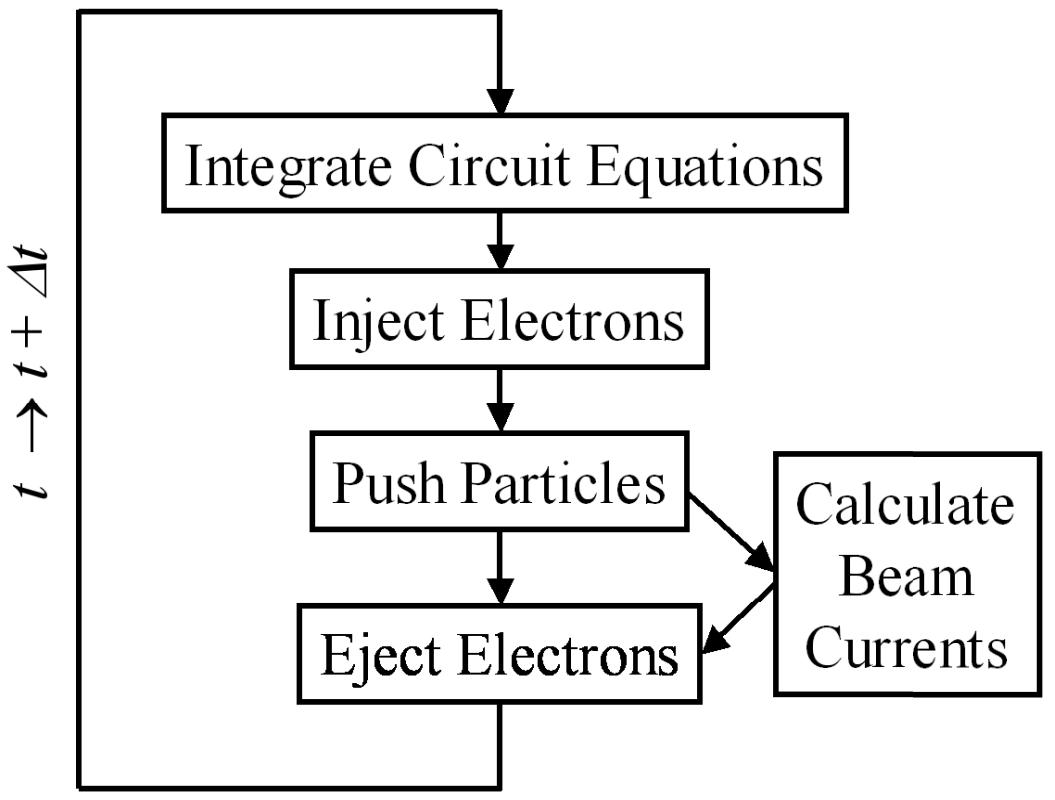

Fig. 2: Flow chart of the simulation procedure [1].

To simulate a three-dimensional Poisson solver for a multiple beam device, and improve accuracy and efficiency, each beam must propagate in its own beam tunnel. Each beam includes its own interaction gap. The Poisson solver, therefore, must include the region within each beam tunnel as well as the overall cavity to include the fields which the beams would experience in the gaps. The accuracy improvement is achieved because a Dirichlet boundary condition can be prescribed for each beam tunnel. In addition, if an inner conductor is automatically generated, then the Dirichlet boundary condition can be prescribed to this inner conductor which improves accuracy for a given number of radial nodes. The efficiency is improved by not having to deal with the boundary condition of the cavity axis.

We use an algorithm to generate the mesh of the beam tunnels for one cavity. The algorithm can easily be extended for multiple cavities along the main axis. The beam tunnels are assumed to have the same (minor) radius $r_b$, and their centers span equal angle distance and on the same (major) radius $R_b$ from the cavity axis. The algorithm takes advantage of the structure of the finite difference mesh to construct the mesh of the beam tunnels only once. At every axial ($z$) node, the same indices of the ($r$, $\theta$) nodes prescribing the beam tunnels are repeated.

Let $R_i$ be the radius of the inner conductor, $N_r$ the number of $r$ nodes from the inner conductor to the outer conductor, $N_\theta$ the number of angles, $d_r$ the mesh $r$-size, $d_\theta$ the mesh $\theta$–size, and $N_b$ the number of beam tunnels. In addition, the number of local angles $n_\theta$ is also needed to construct each tunnel. It's noted that $n_\theta << N_\theta$ for an accurate representation of the tunnels. We assume the first tunnel is on the $x$ axis. The pseudo algorithm is as follows.

The local angle distance is $\delta_\theta = 2\pi/n_\theta$. For each tunnel $m$ in $[0,N_b]$ set the global angle of the tunnel center $\theta_m = md_\theta$. Then for each local angle $n$ in $[0,n_\theta]$

1. Set the local angle in the tunnel $\theta_n = n\delta_\theta$,
2. Construct the local coordinates ($r_l$,$\theta_l$) on the tunnel radius using ($r_b$,$\theta_n$),
3. Transform cylindrical coordinates from local ($r_l$,$\theta_l$) to global ($r$,$\theta$) using ($R_b$,$\theta_m$),
4. Search for global mesh indices ($i$,$j$) of ($r$,$\theta$) space using $R_i$, $d_r$, $d_\theta$,
5. Obtain one node on the tunnel in the global mesh, from the global mesh indices ($i$,$j$) and (6) store the pair ($i$,$j$) for later boundary condition assignment.
6. Finally, apply the Dirichlet boundary condition to each mesh node on the tunnels along the cavity $z$ axis except for the tunnel gap. The axial location of the gaps is preset.

The three-dimensional Poisson solver presents the single greatest computational load in NEMESIS. As a result, we have parallelized the three-dimensional Petsc solver using the Message Passing Interface (MPI). Running on a Windows laptop computer with four cores, an example of the improvements in the overall runtime is shown in Fig. 3 where we plot the normalized run time versus the number of cores for the simulations discussed below. The runtime decreased by a factor of more than three by engaging all four cores.

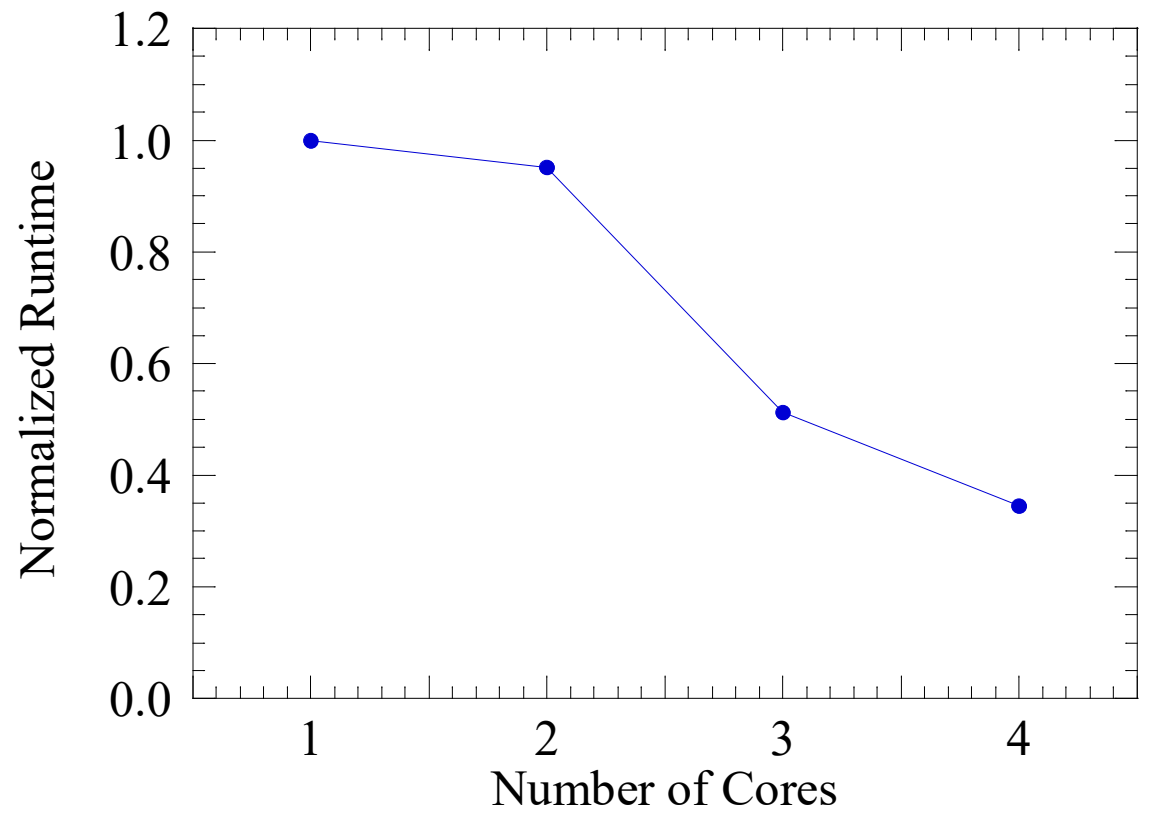

Fig. 3: Normalized runtime for NEMESIS using the parallelized three-dimensional Poisson solver.

The inclusion of the 3rd harmonic is done via the following model for the drive current, where $I_p$ is the peak current, $\varepsilon$ denotes the ratio of the 3rd harmonic to the fundamental and $\varphi$ is the phase shift of the 3rd harmonic relative to the fundamental.

$$I(t) = I_p \left[ \sin\left( \pi \frac{t}{\tau_{width}} \right) + \varepsilon \sin\left( 3\pi \frac{t}{\tau_{width}} + \varphi \right) \right]^2 , \quad (1)$$

for $t < \tau_{width} < 1/f$ and zero otherwise, where $f$ is the wave frequency and $\tau_{width}$ denotes the portion of the wave period over which electrons are drawn off the grid. The ratio of the average to peak current using this model for the drive current is

$$\frac{I_{avg}}{I_p} = \frac{1+\varepsilon^2}{2} \frac{\tau_{width}}{\tau_{period}} . \quad (2)$$

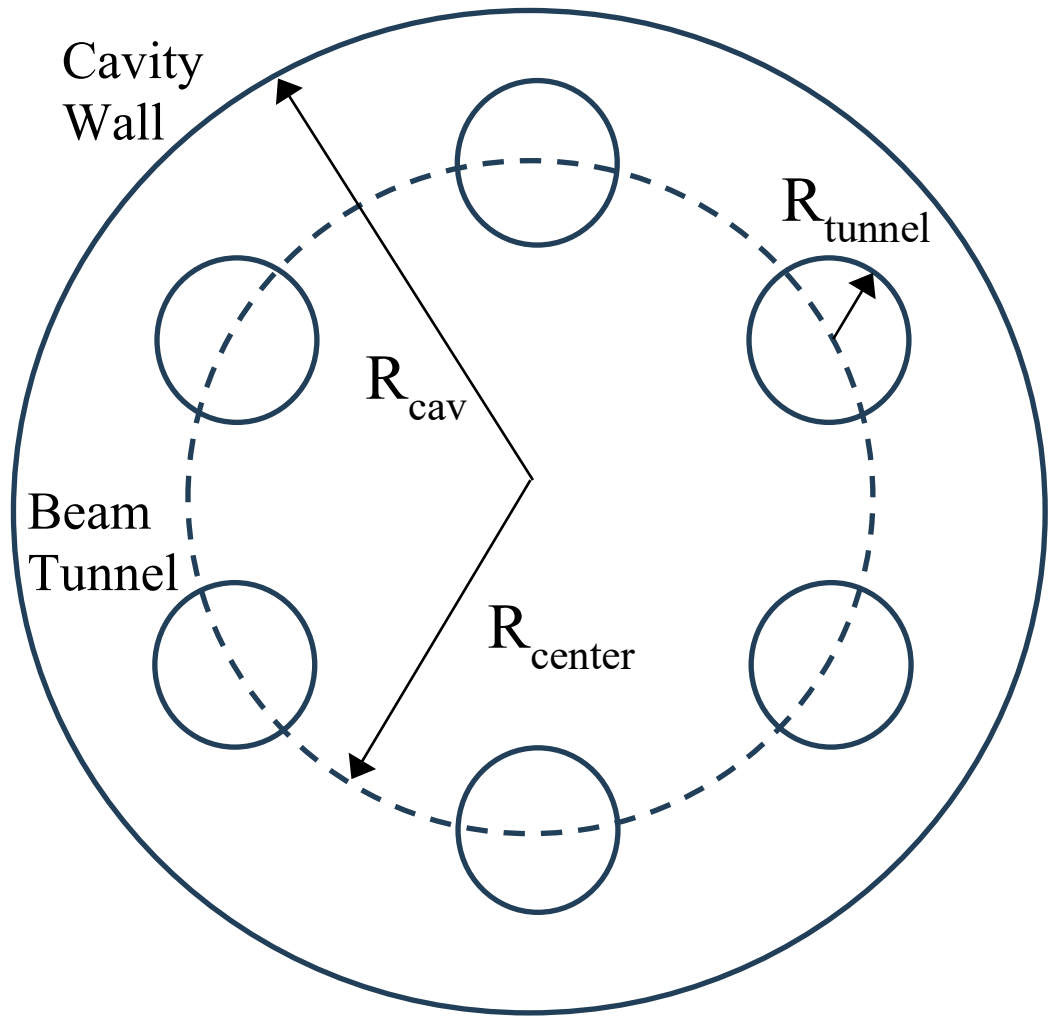

Fig. 4: Schematic illustration the cross section of a 6-beam MBIOT.

Fig. 5: Schematic illustration of a side view of the multi-

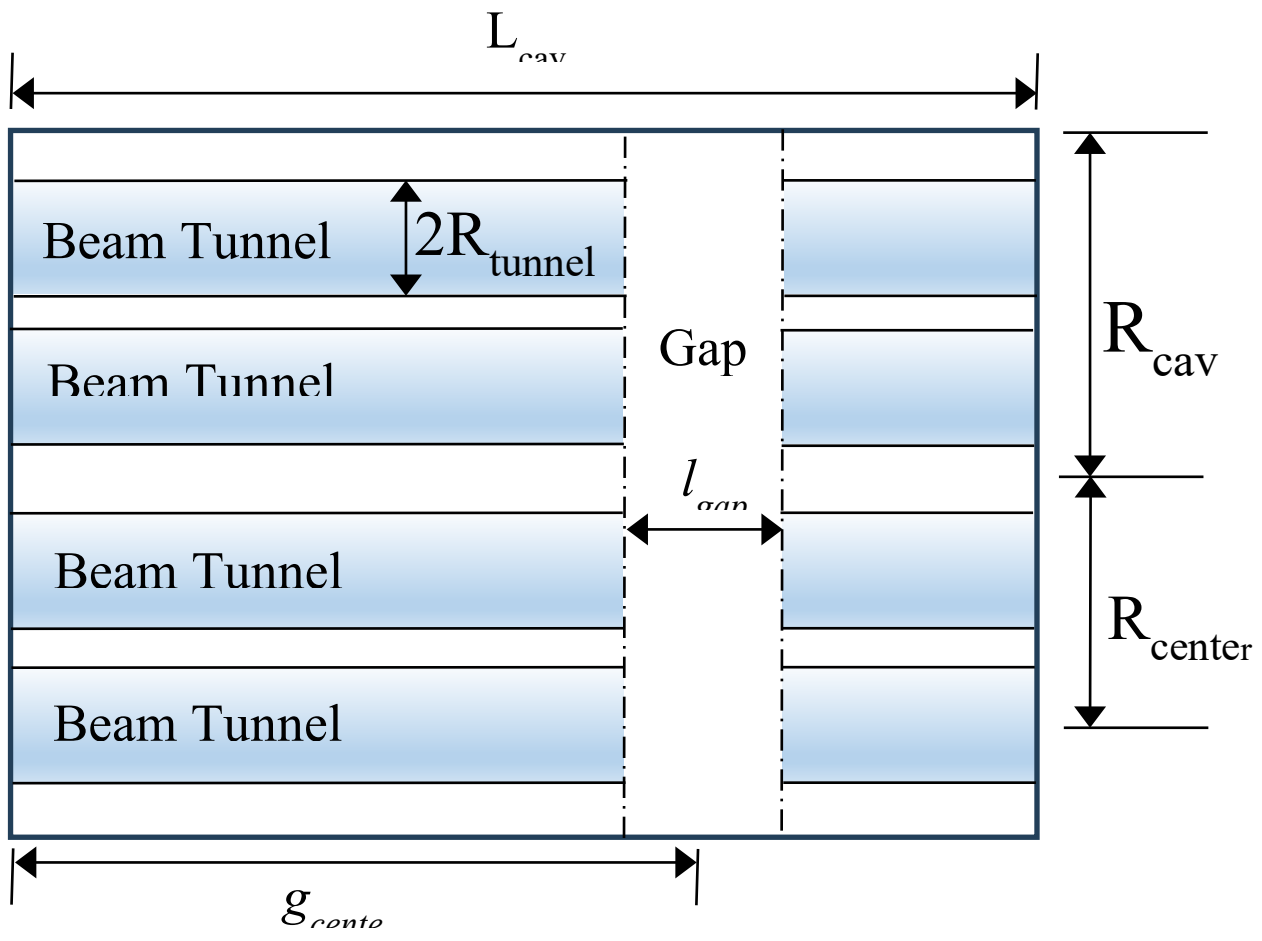

beam cavity showing individual beam tunnels.

We consider an 8-beam MBIOT. A schematic illustration of the cross section of an MBIOT with six beams is shown in Fig. 4 for clarity, and a side view of the cavity is shown in Fig. 5.

We refer to these simulations as whole-cavity simulations because we include all the beams in the simulation including when they are within the beam tunnels or in the gaps. This contrasts with many simulations of multi-beam IOTs and klystrons in which only a single beam is simulated and the result is then scaled by the number of beams.

The configuration consists of a multi-beam cavity tuned to 700 MHz with $R/Q = 80\ \Omega$ and a loaded $Q = 40$ with a radius of about 11.0 cm, a length of 9.144 cm, and with the gap center located 4.571 cm downstream from the entrance to the cavity. The optimal gap length and the optimal position of the gap center were found through a multiple-parameter series of simulations. The optimal gap length was found to be about 3.9 cm. The electron beam voltage and average current are 35.155 kV and 7.25 A (over the eight beams) respectively. A solenoidal focusing field is used with an amplitude of 126 G which is close to the Brillouin field when the electron cyclotron frequency $\Omega_e = \sqrt{2}\omega_{pe}$, where $\omega_{pe}$ is the electron plasma frequency [15]. The ratio of the average to peak current is 0.15. In the absence of any 3rd harmonic drive, this implies that the ratio of the width of the pulse to the resonant period (= $1/f$, where $f = 700$ GHz) is 0.30. Note that from Eq. (2) the ratio of the average to peak current depends upon the resonant period, the bunch width and the relative strength of the 3rd harmonic. In this work we held $I_{avg}/I_{peak}$ and the resonant period fixed so that the bunch width varied with the relative strength of the 3rd harmonic.

The variation in the performance versus the cavity radius ($R_{cav}$) is shown in Fig. 6 with (blue) and without (red) the 3rd harmonic drive for a cathode radius ($R_{cathode}$) of 1.05 cm and a beam tunnel radius ($R_{beamtun}$) of 1.60 cm. Here we observe that the performance decreases with increasing cavity radius and we focus on a cavity radius of 11.0 cm.

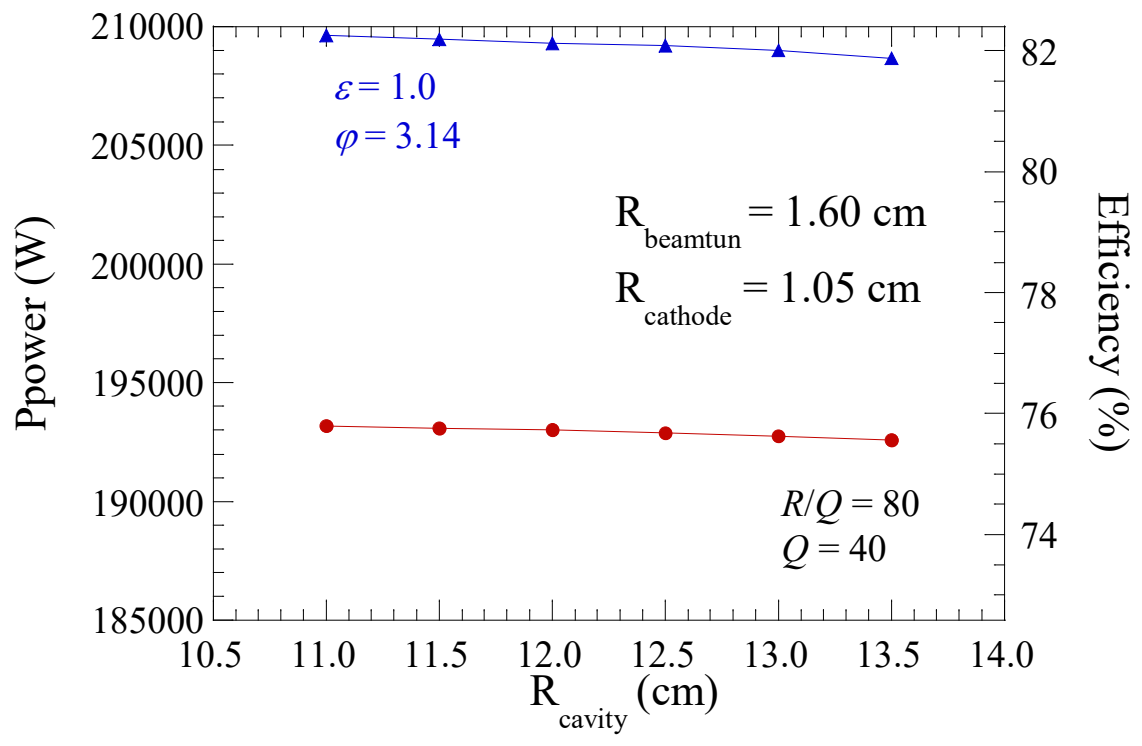

Fig. 6: Variation in the output power and efficiency as functions of the cavity radius. The blue (red) line corresponds to the case with (without) the 3rd harmonic drive.

The variation in the performance versus the beam tunnel radius is shown in Fig. 7 both with (blue) and without (red) the 3rd harmonic drive and for a cavity radius of 11.0 cm and a cathode radius of 0.975 cm. It is clear from the figure that the performance peaks for a beam tunnel radius of about 1.60 cm (as used in Fig. 6). Hence, we will focus on this beam tunnel radius.

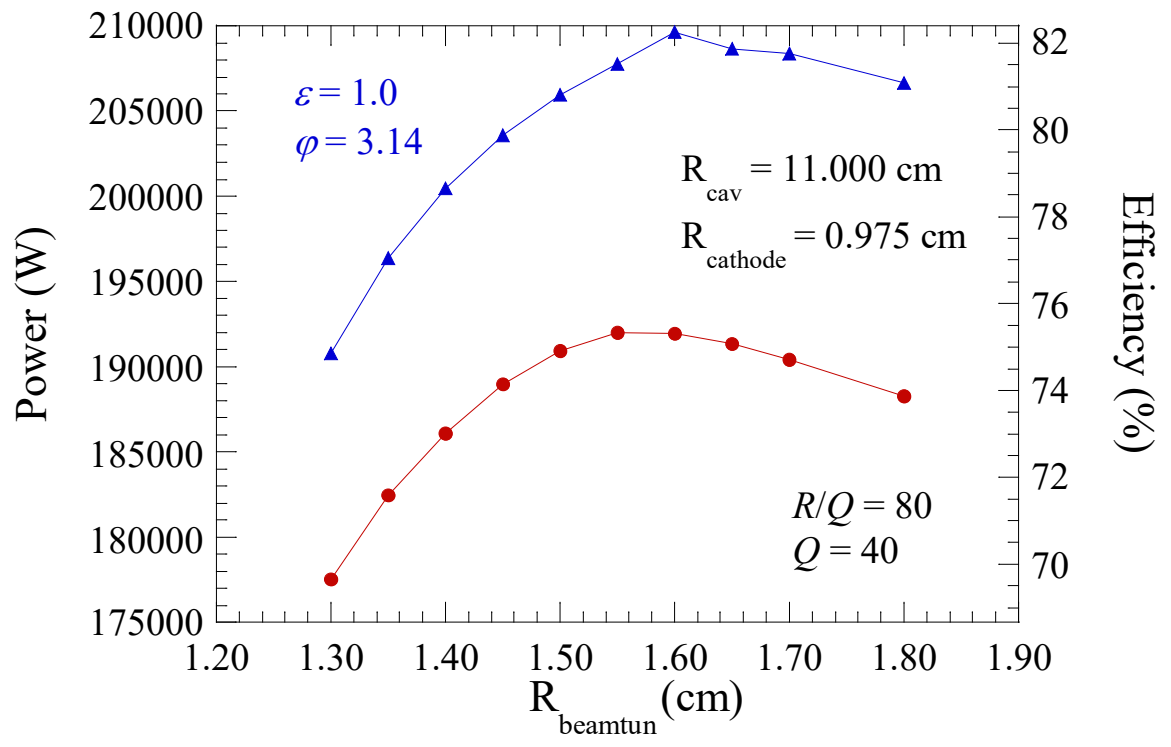

Fig. 7: Variation in the output power and efficiency as functions of the beam tunnel radius. The blue (red) line corresponds to the case with (without) the 3rd harmonic drive.

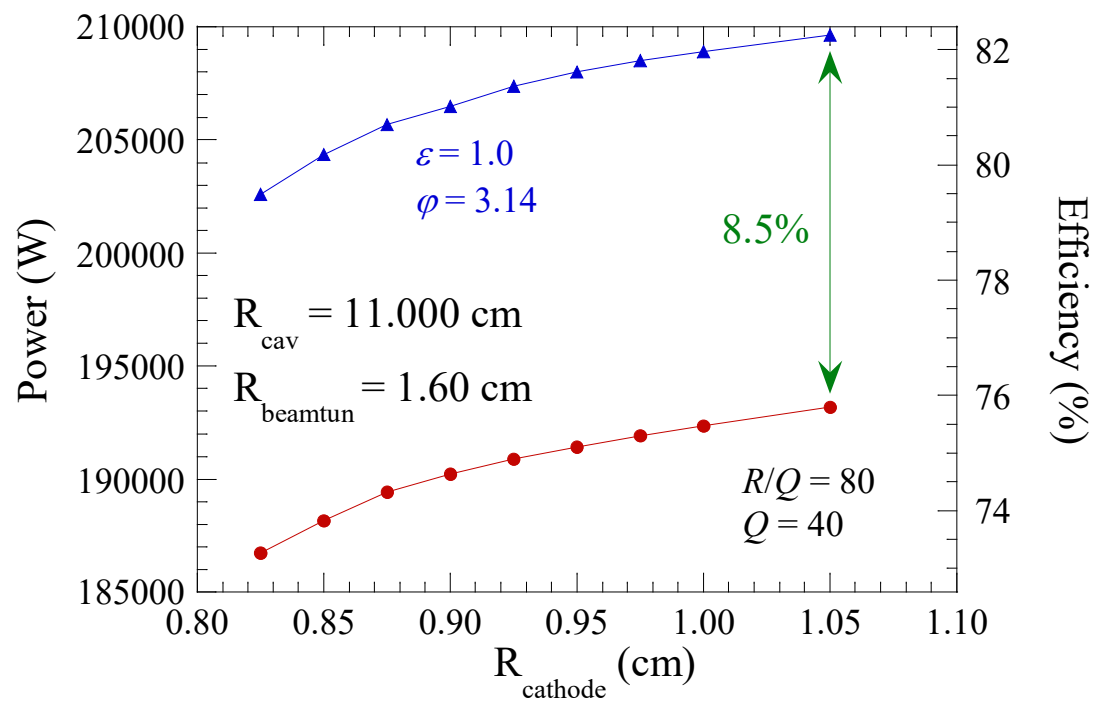

Fig. 8: Variation in the output power and efficiency as functions of the cathode radius. The blue (red) line corresponds to the case with (without) the 3rd harmonic drive.

We now consider the performance as a function of the cathode radius using the optimized values of the cavity radius (11.0 cm) and beam tunnel radius (1.60 cm). The variation in the output power and efficiency versus the cathode radius is shown in Fig. 8 for both the absence (red) and presence of the 3rd harmonic drive (blue). We observe that the use of the 3rd harmonic drive results in an increase in power of about 8.5%. The performance improves as the cathode radius is increased up to a cathode radius if 1.05 cm which is the largest radius that we consider for the MBIOT design. Overall efficiency of about 82% is indicated at this cathode radius when the 3rd harmonic is applied to the drive.

## III. MBIOT COMPONENT DESIGN

In this section we discuss three elements of the MBIOT design: the grid, the input coupler and the output cavity.

### A. The Grids

A program goal was to replace the Pyrolytic Graphite (PG) grid in the IOT by a molybdenum (moly) grid, which is much easier to fabricate, and still preserves the existing beam optics. A moly grid will also operate at a lower temperature due to its lower emissivity and provide a higher design safety factor. However, it has a higher thermal expansion than PG; therefore, a detailed thermal and structural analysis of the electron gun was needed to ensure its emitter-grid spacing at temperature, and beam optics of the PG gridded and moly gridded gun were comparable.

Simulating a PG grid is problematic because of its orthotropic properties and spherical shape. Its thermal conductivity in the *ab* plane is two orders of magnitude larger than in the *c* plane. Its thermal expansion in the *ab* plane is almost one order of magnitude larger than in the *c* plane. Orthotropic materials require the ability to prescribe material coordinates for the orthotropic parts in the model, so that proper coordinate transformation can be performed in the finite element formulation. Unfortunately, neither Ansys nor SolidWorks simulation provides the option to prescribe a spherical coordinate system. They permit only Cartesian and cylindrical coordinates. If either Ansys or SolidWorks were to allow spherical coordinates, we will just need to segment the existing grid into two parts, one for the flat rim with a Cartesian coordinates, and one for the spherical section with the spherical coordinates. Due to this shortcoming, the PG grid is analyzed with two approximations: (a) Orthotropic PG with cylindrical coordinates (b) Isotropic PG with *ab* plane properties.

Figure 9 shows the model of the grid with its support assembly. This model provides the grid displacement only. Thus, for the sake of hot spacing comparison between PG and moly grids, one can think of a nondisplaced emitter as the referenced position. For the heat transfer model, the copper stem is air cooled, the grid top radiates to a heat sink kept at 150°C, and the grid bottom is radiated by the emitter kept at 1050ºC and loaded by the electron beam of 2.5 W. Air flow rate can be adjusted to obtain the same stem temperature when the grid material is changed to moly. For the mechanical model, the bottom of the stem is kept fixed, and the grid rim is assumed to bond to the grid deck. Other part interfaces are also assumed to be bonded.

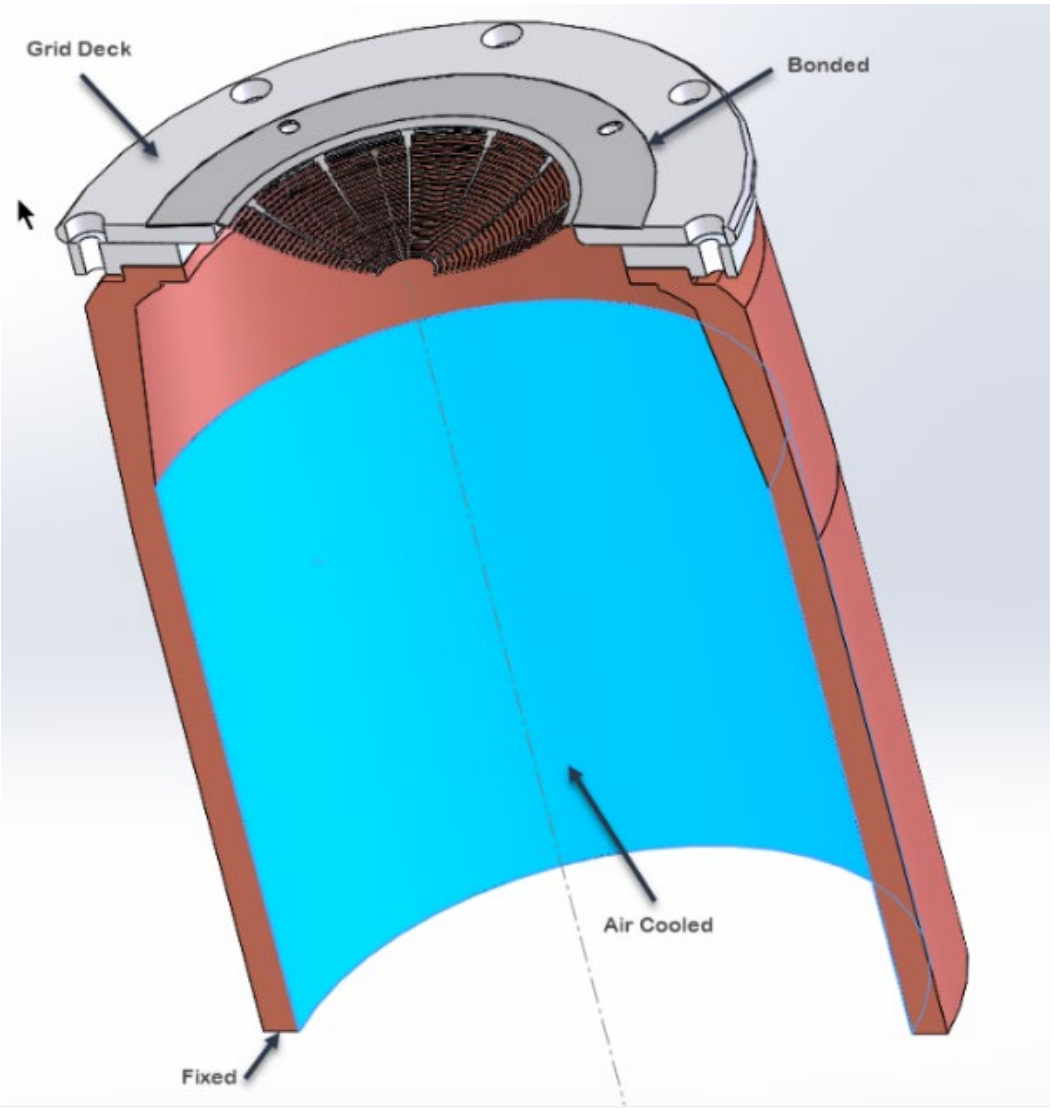

Fig. 9: CAD model of the grid and supports.

With approximated orthotropic PG, for which the material, cylindrical coordinates were applied to the grid, the c plane properties will have larger effects making the grid less conductive and lower thermal expansion along the lateral surface. Temperature would be more uniform in the spherical section; consequently, lower compression and lower axial displacement than those of approximated isotropic PG. Due to the thinness of the grid, the isotropic approximation of pyrolytic graphite was more realistic than the actual orthotropic properties with *incorrect* material coordinates particularly for thermal calculation. However, the isotropic formulation required positive Poisson's ratio that the PG *ab* plane did not. This was very likely leading to inaccurate stress and displacement.

Results of the analysis for the model shown in Figure 9 are summarized in Table 1. It shows the orthotropic PG grid temperature being the highest at 886ºC, and isotropic PG at 568ºC.

| Material | Max Temp (°C) | Tensile Strength (MPa) | Stress (MPa) | Rim Displ (µm) | Center Displ (µm) |
|---|---|---|---|---|---|
| Ortho PG | 886 | 124 | 38.3 | 45.3 | 40.5 |
| Iso PG | 560 | 124 | 62.8 | 51.8 | 78.0 |
| Moly | 472 | 430 | 215 | 43.5 | 0.52 |

Table 1: Grid thermal and mechanical characteristics. The emittance structure is assumed to be fixed with zero displacement for comparison. The stress is calculated at the screw holes.

As mentioned above, the isotropic PG temperature at 560ºC was very likely more realistic and comparable to the moly grid temperature at 472ºC. The center of the PG grid moving away from a *fixed* emitter was in the range of 40.5 µm to 78.0 µm, which was two orders of magnitude greater than that of the moly grid at 0.52 µm.

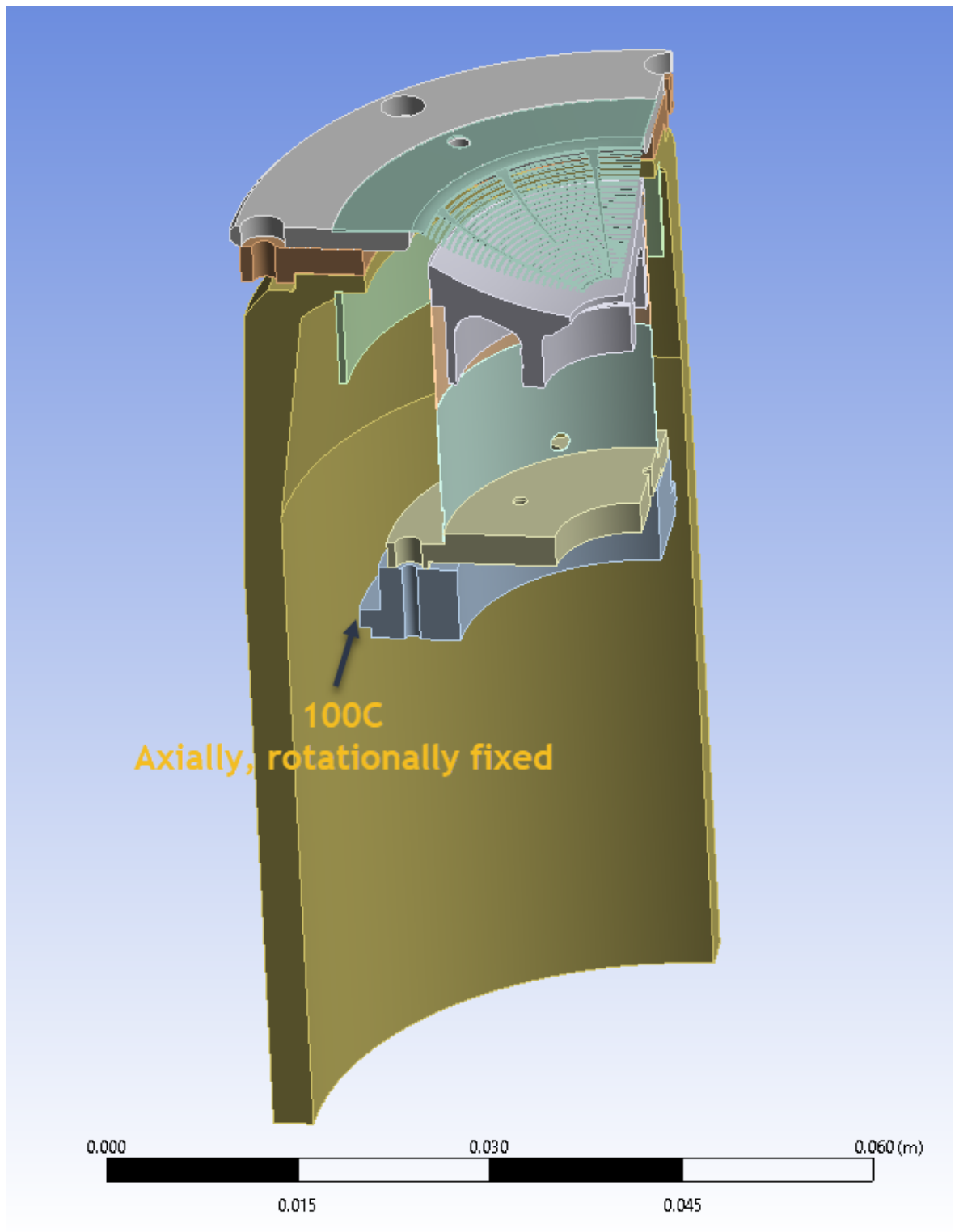

Fig. 10: Ansys model of the moly gridded gun. Only a quarter of the CAD model is used.

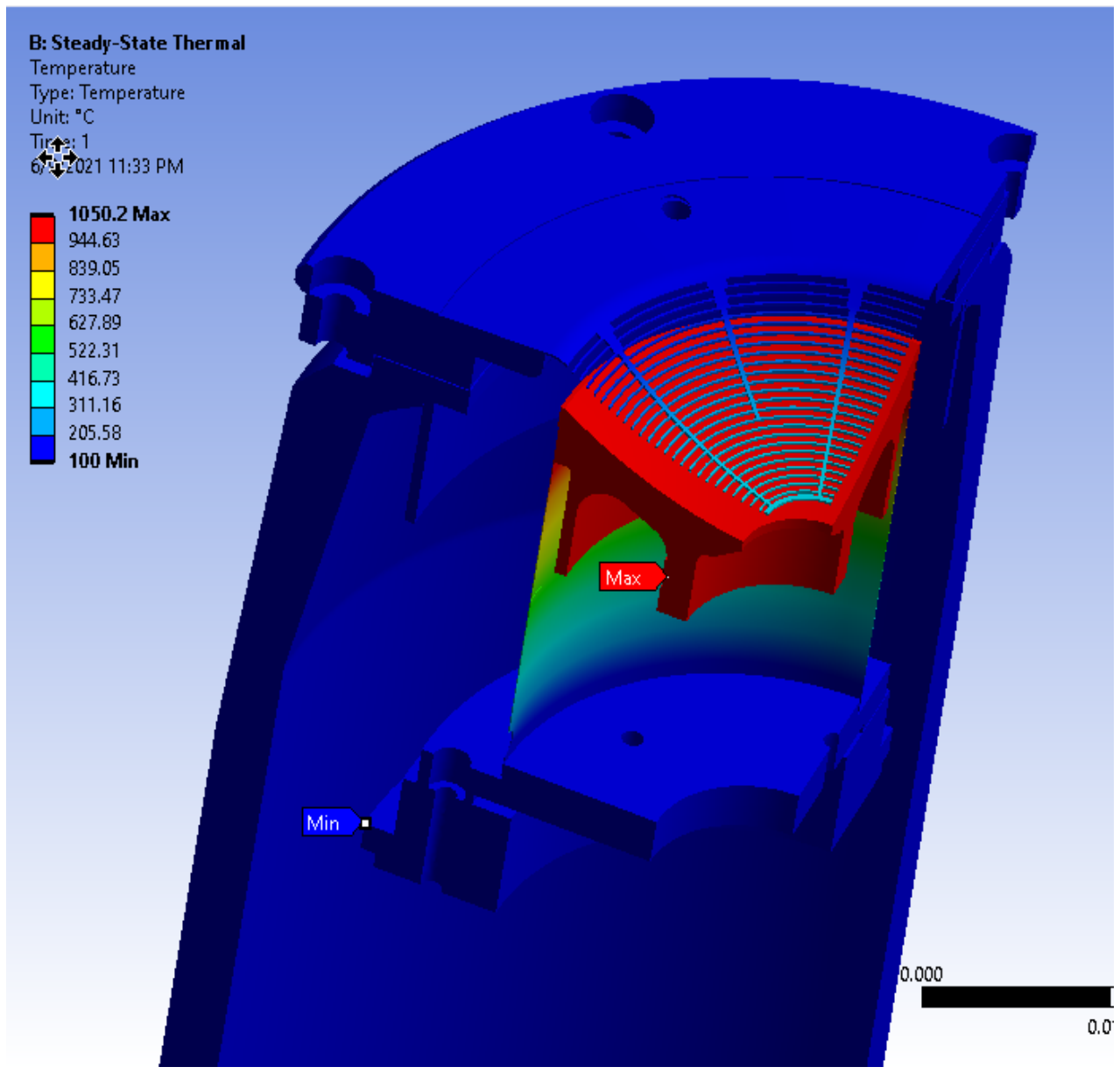

Fig. 11: Temperature of the moly gridded gun.

Since it was difficult to accurately obtain the hot spacing of the PG grid to use as a benchmark design, we decided to pursue the full analysis of the emitter and grid structures to obtain the emitter-moly-grid spacing at temperature, and use it to verify the beam optics. This model for the moly gridded gun is shown in Fig. 10. The same boundary conditions as shown in Fig. 9 were used in addition to the boundary conditions being imposed at the OD of the emitter support. Only one quarter of the model is used in the simulation to reduce computer time. To further reduce the compute time, the radiation exchanges between the heat shields were analytically derived via circuit representation of the surface and geometrical resistances to scale the surface emissivity and area accordingly. The volumetric heat generated in the emitter was iteratively prescribed to achieve the emitter surface temperature of 1050ºC. The temperature of the moly gridded gun structure is shown in Fig. 11. The corresponding axial displacement is shown in Fig. 12, and its close up in Fig. 13. The emitter and grid OD move closer to each other from their cold positions by 105 µm. The emitter-grid ID spacing was also reduced, but less, by 73.8 µm.

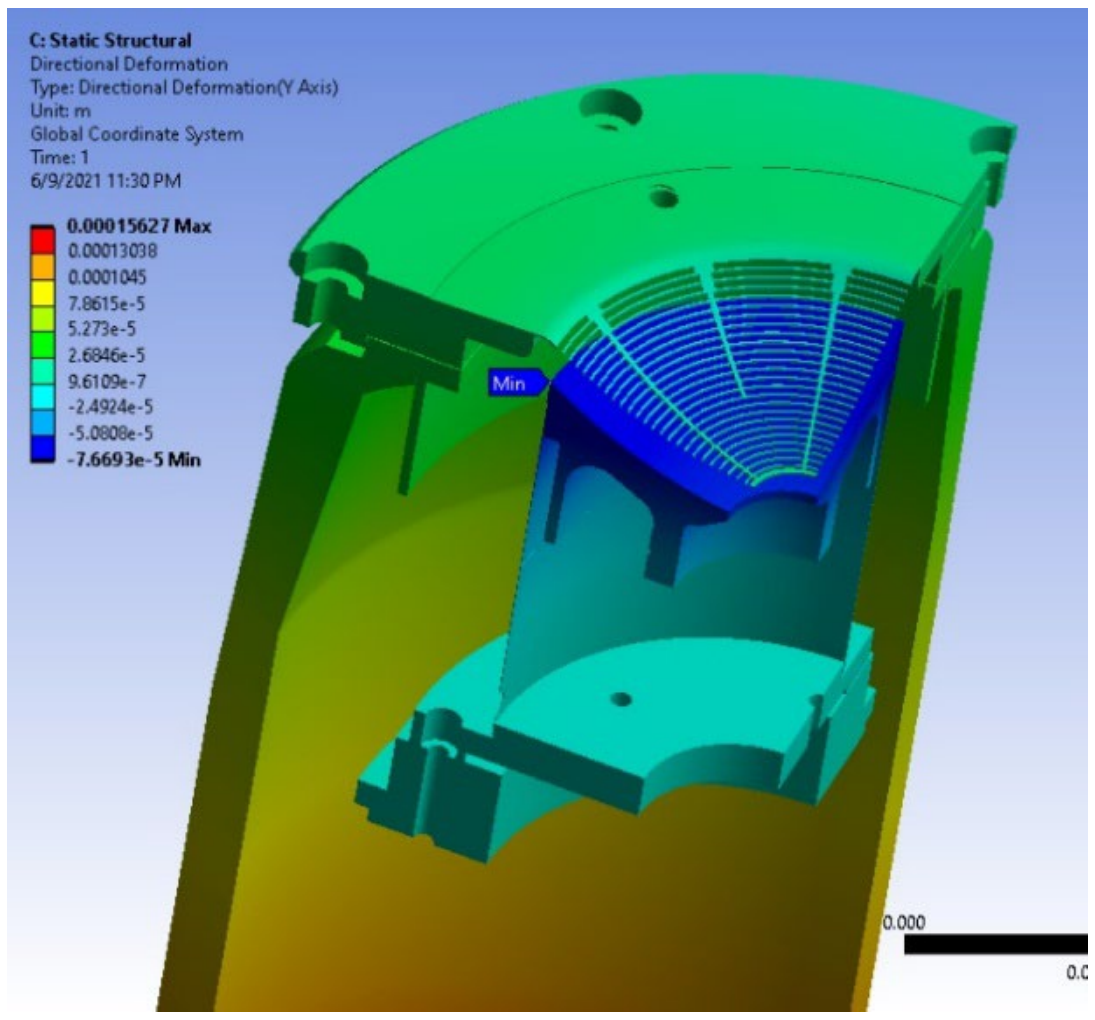

Fig. 12: Axial displacement of the moly gridded gun.

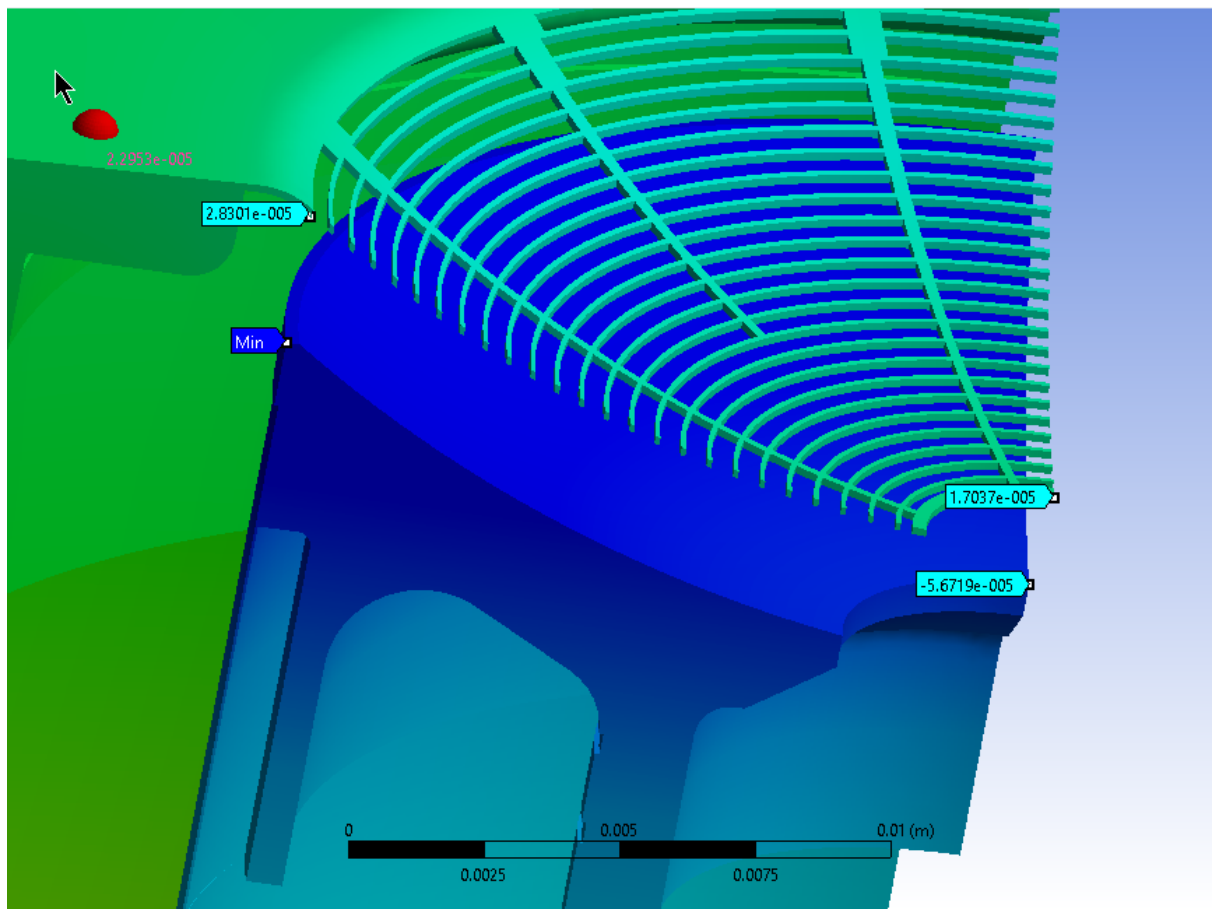

Fig. 13: Close up of the axial displacement of the moly gridded gun.

With the above hot dimension, we performed the beam optics simulation including optimization to achieve the required emission for moly grid.

## B. The Input Coupler

The coupler was designed to simultaneously drive the grids of eight guns at both the fundamental (~700 MHz) and the third harmonic (~2100 MHz). Initially it was hoped that the coupler would combine the frequencies through separate feeds. However, no practical geometry with two feeds was found, and it was decided to combine the feeds with an external coaxial hybrid T.

The design of the coupler was complicated by the basic fact that the roots of the eigenfunctions of a circular cavity – Bessel functions – are not uniformly spaced. Thus, it is not possible with a simple circular cavity to have both the fundamental and third harmonic resonant at the same frequency. With a 700 MHz fundamental in the $TM_{01}$ mode, the $TM_{03}$ mode will have a resonance at 2515 MHz.

This required the introduction of a structure to modify the modes. The chosen arrangement was an annulus at the minimum of the third harmonic to primarily influence the fundamental. The geometry is shown in Fig. 14. There are eight electron guns evenly arranged at a radius of 90.24 mm (3.560 in). The gun grids are connected to the cavity by short coaxial lines, the inner conductors of which are hollow and terminate at the outer cavity wall. This allows for the heater leads and the air outlet of the grid cooling. The design also includes tuners to adjust both resonances, allowing for the cavity to be used over a range of frequencies.

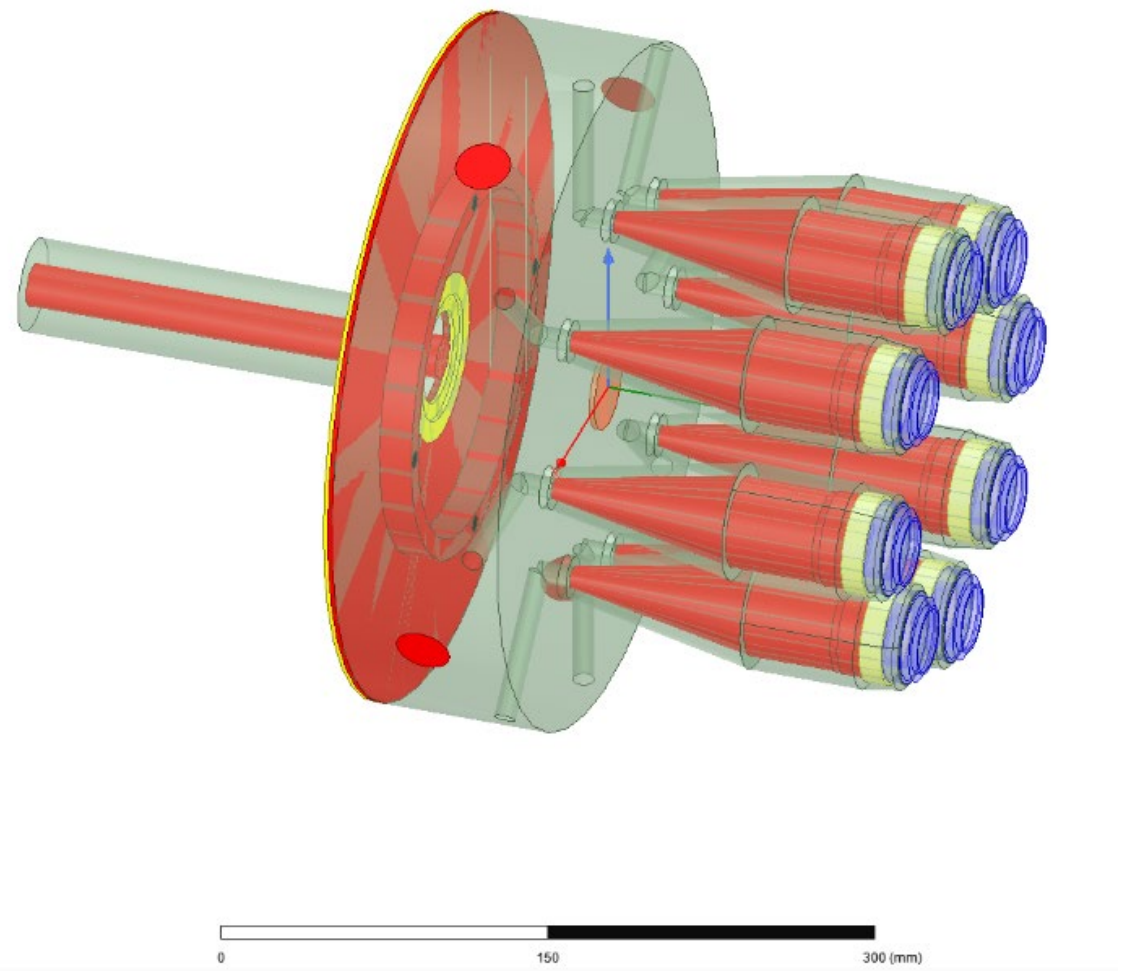

Fig. 14: Geometry of the input cavity.

The result is a tunable cavity that provides the required grid voltage (~62 V) with an input of 4 kW at the fundamental. This implies a gain of 20 dB. But a hybrid T in the input coax will reduce that gain by about 4 dB, so the net gain will be about 16 dB. For the same input power, the voltage at the third harmonic will be about 35 V.

The cavity was simulated using ANSYS HFSS. As shown in Fig. 14, the model included the actual coupling cavity and the electron gun grid-driving structure. The grid was simulated using a lumped-load with an impedance provided by the gun manufacturer.

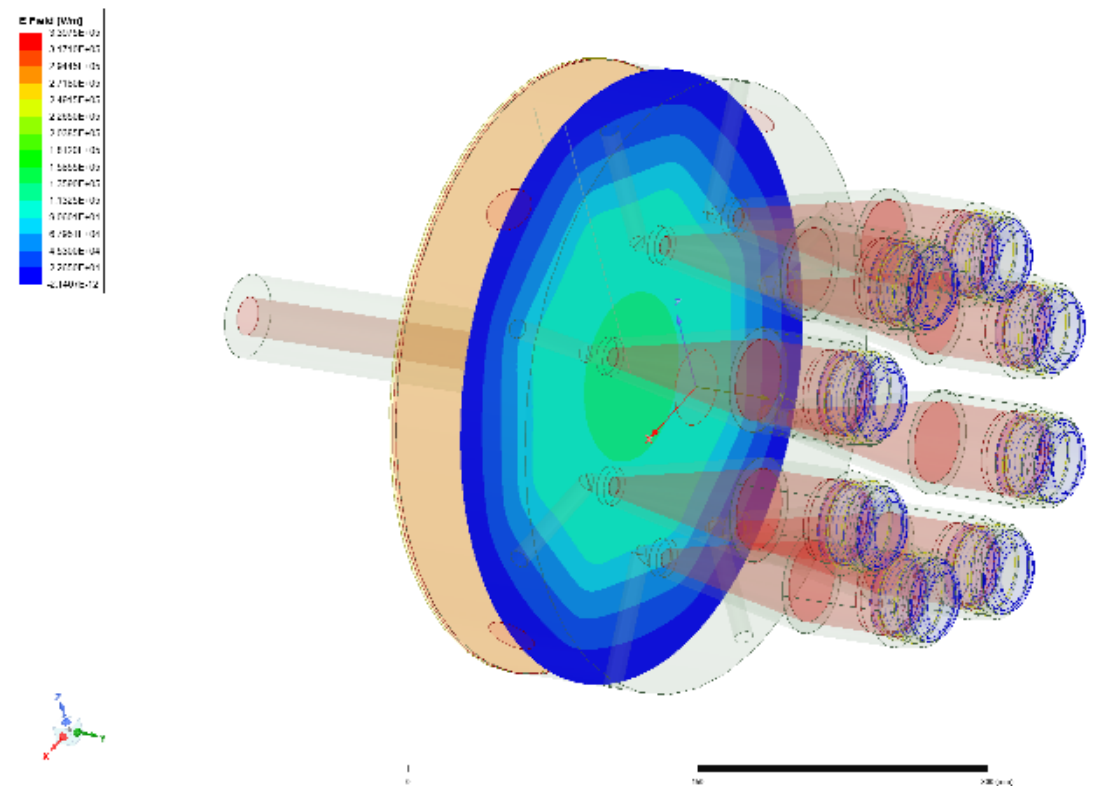

Fig. 15: Fundamental on the midplane of the cavity.

At resonance, the complex magnitude of the RF electric fields is given in Fig. 15 and Fig. 16. As expected, (and desired) the fields are azimuthally symmetric.

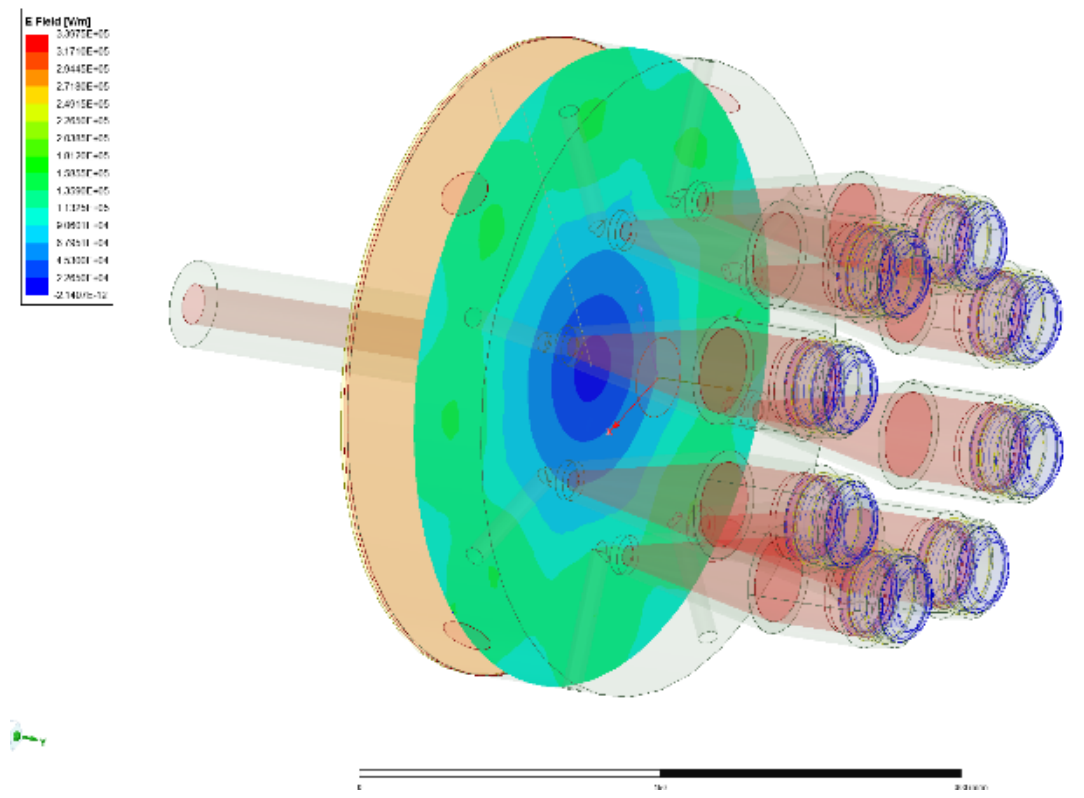

Fig. 16: Third harmonic on the midplane of the cavity.

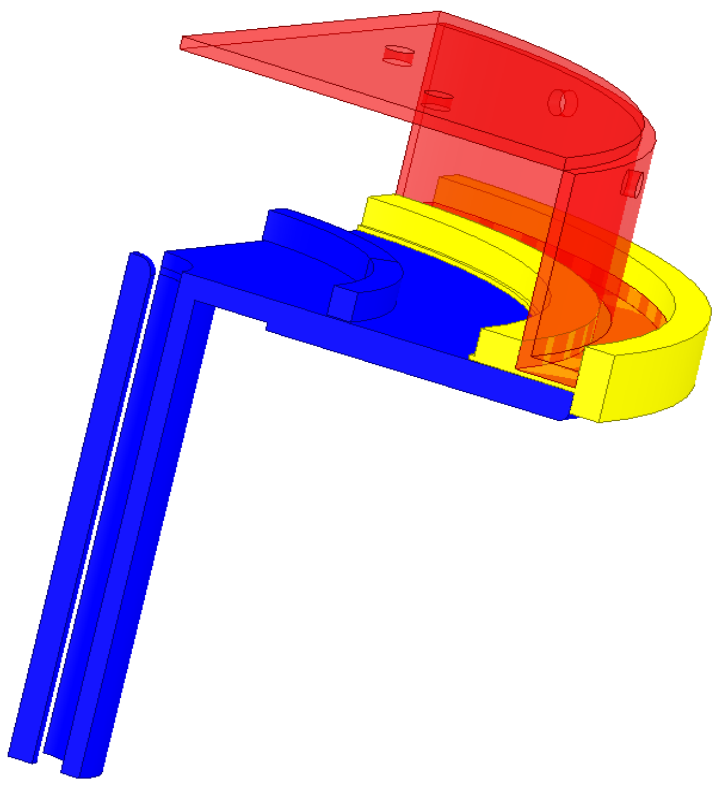

Fig. 17: Quarter model of the approach for the isolation. The insulator is shown in yellow. The bottom plate (blue) is at ground potential while the top (red) is at the cathode potential.

The output of the cavity is at the cathode potential. High voltage isolation of the input was provided by including insulators as shown in Fig. 17.

With the insulator included, a scan was made of the resonant frequency and grid voltage as a function of the cavity height and top and side tuners. A cavity radius of 172 mm gave a tuning that is close to being centered on 700 MHz. The results of scanning the side and top tuners are shown in Fig 18. With adjustment of both tuners, simultaneous resonances at both the fundamental and 3rd harmonic can be achieved from 698 MHz to 703 MHz.

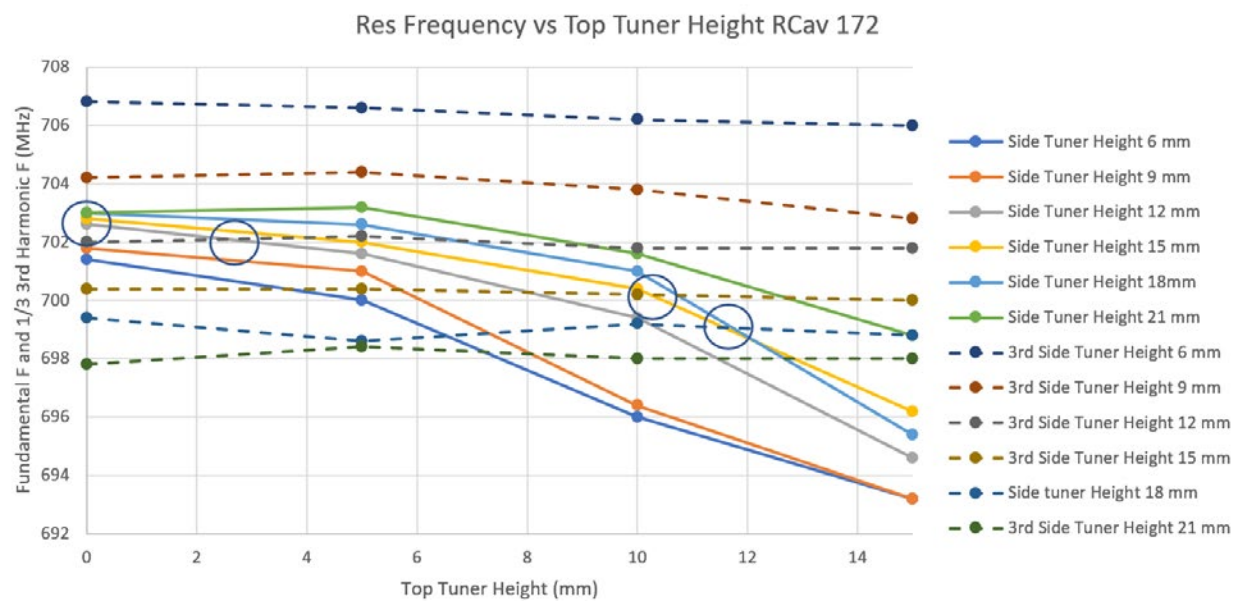

Fig. 18: Plot of the resonant frequencies as a function of the top tuner height for several side tuner heights. The cavity radius is 172 mm. Intersections of the fundamental and the (dashed) $3^{rd}$ harmonic/3 lines give the operating frequencies.

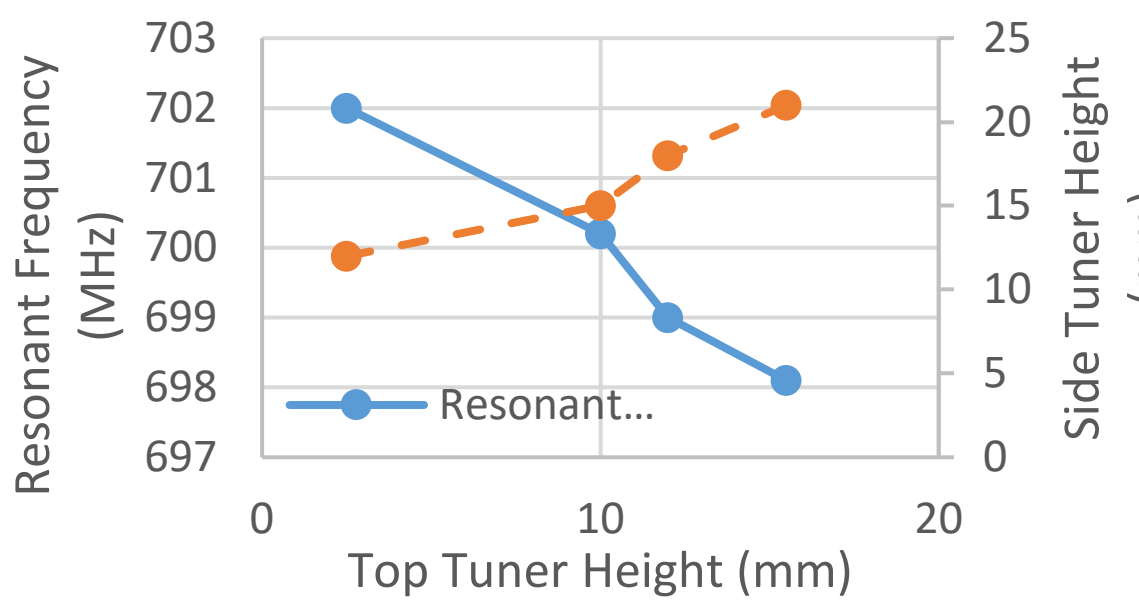

Fig. 19: Plot of the resonant frequencies of the points of intersection from Fig. 18.

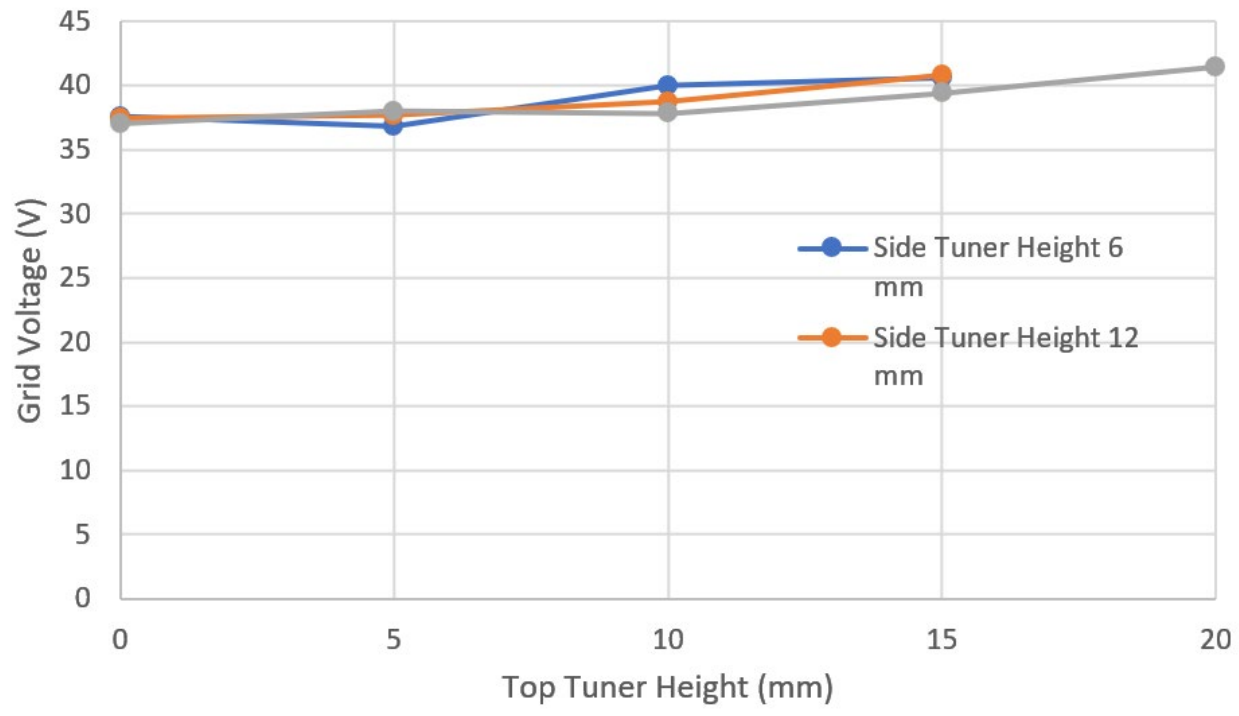

Fig. 20: Gun grid voltage with the cavity driven with 2 kW at the fundamental.

Figure 19 shows the intersection frequencies. Figures 20 and 21 show the grid voltages for the parameter space. The voltage at the fundamental is almost independent of the tuner heights. The third harmonic voltage is weakly dependent on the top tuner height, but has a significant dependence on the side tuner height. An input power of 4 kW at the fundamental will provide about 61 V required to fully modulate the electron guns. The same power at the third harmonic will produce between 30 V and 45 V.

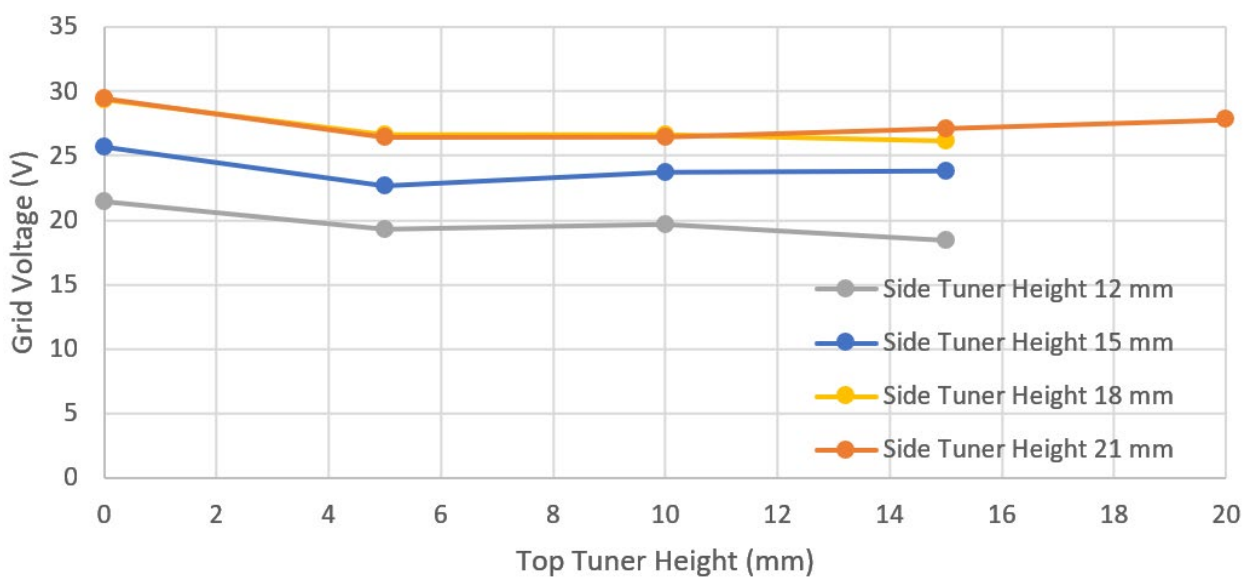

Fig 21. Gun grid voltage with the cavity driven with 2 kW at the third harmonic.

To summarize, a resonant structure has been designed to provide coupling from a single input to the eight electron gun grids. With appropriate adjustment of tuners, the cavity can be tuned from 698 MHz to 702 MHz. Four kilowatts is required to drive the gun grids at the fundamental.

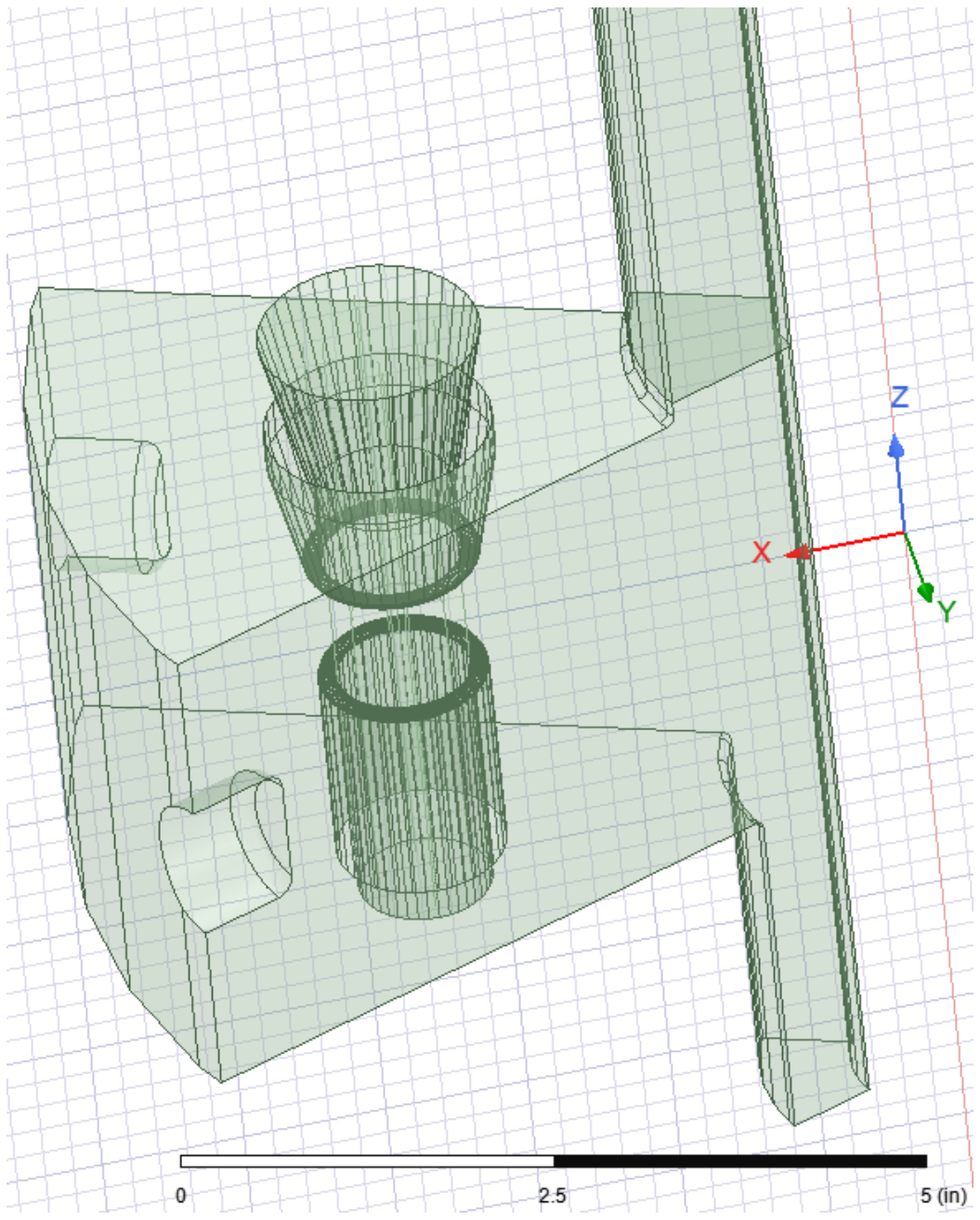

Fig. 22: HFSS model of a 1/8th slice of the Output Cavity.

## C. The Output Cavity

For the output cavity we chose a coaxial cavity design operating in the fundamental mode. A coaxial cavity design allows flexibility in the size of the "bolt circle" where the beams are placed and optimization of the electric field (amplitude and uniformity) in the drift tube gap.

The RF Design of the eight-beam output cavity has been created using HFSS. It features eight reentrant drift tubes positioned in a circle. The RF power is extracted from the output cavity through a center coaxial line. Fig. 22 shows the HFSS model of a 1/8$^{th}$ slice of the output cavity. The beam tunnel is in the middle of the slice. The cavity has eight tuning posts protruding from the outer cavity wall for adjusting the resonant frequency in cold test after brazing the cavity. In Fig. 22 the tuning posts are sectioned by the symmetry planes.

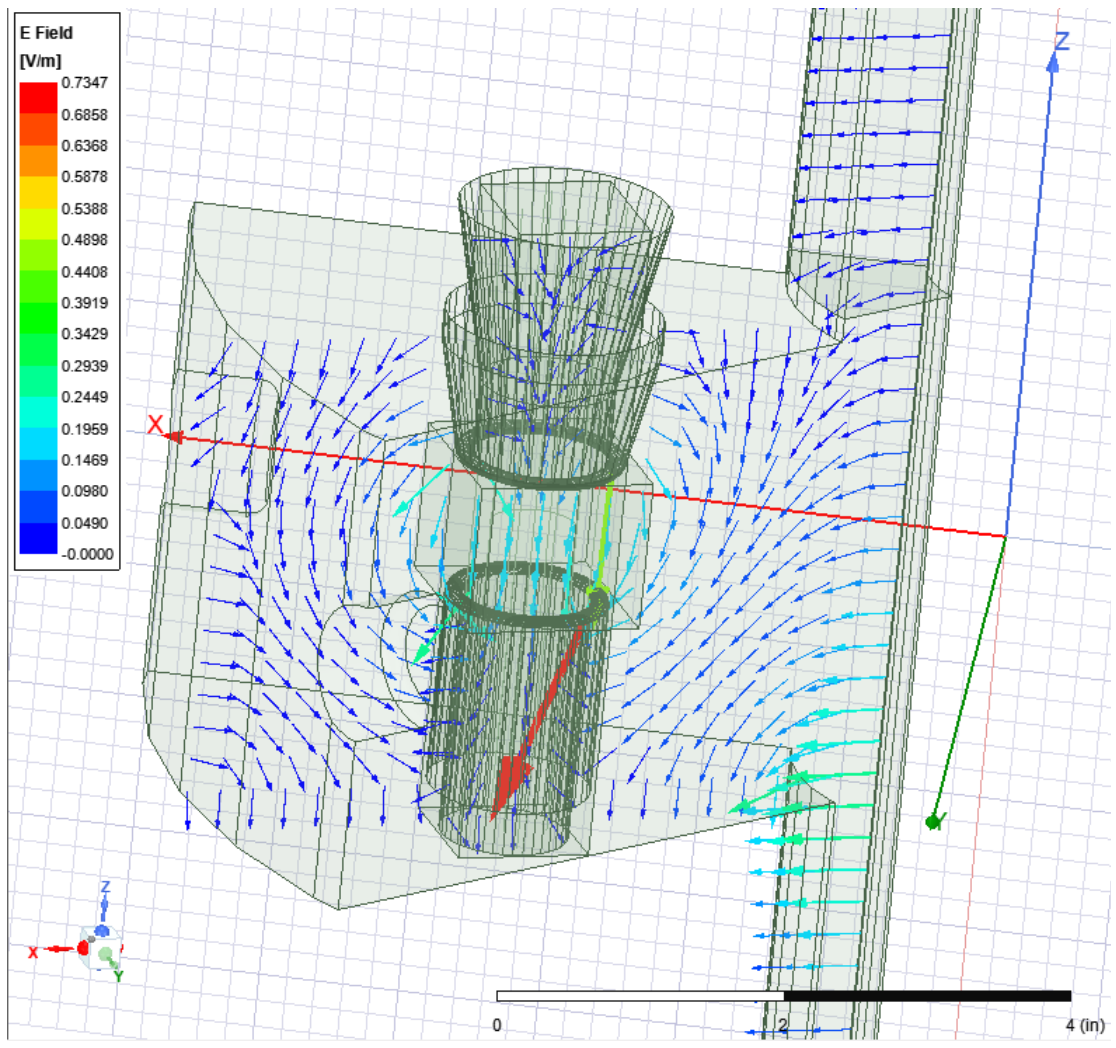

Fig. 23:HFSS model of a 1/8th slice of the Output Cavity. The normalized electric field vectors are shown on a radial plane through the beam tunnel.

The resonant frequency of the output cavity is 700 MHz, the external Q is 120, and the R/Q in the beam center is 185 Ω (R/Q related to the stored energy in 1/8$^{th}$ of the cavity volume). Figure 23 shows the HFSS model including the electric field vectors on a radial plane through the beam tunnel.

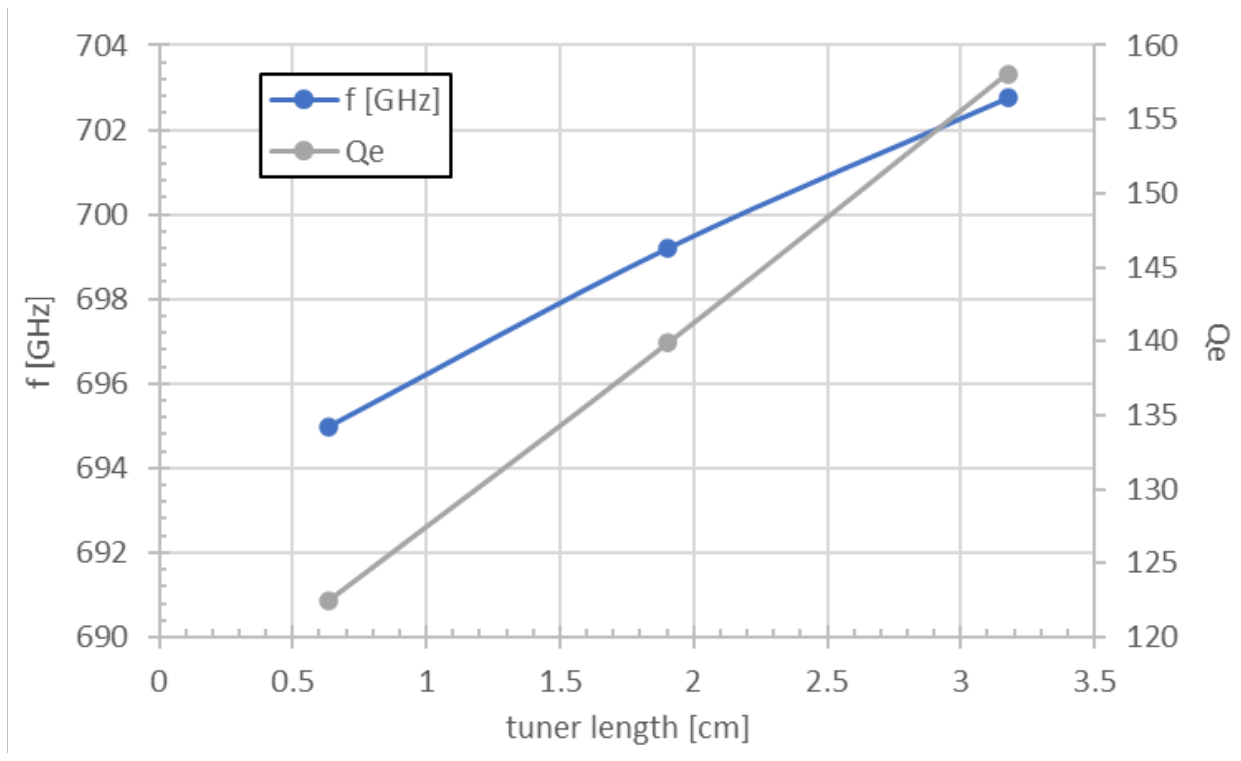

Fig. 24: Simulated output cavity frequency and external Q versus length of the tuning stubs.

The main adjustment of the cavity resonant frequency will be accomplished by machining the cavity diameter. The fine-adjustment of the cavity frequency will be done in cold test after brazing the cavity. Figure 24 shows how the resonant frequency can be adjusted through the depth of the tuning posts. The adjustment will also impact the external Q.

The external Q of the output cavity will then be adjusted through the length of the coax short, as shown in Fig. 25. While all cavity parameters are influenced by the various cold-tuning features, the frequency dependence on the length of the coaxial short is small and has a negative slope. This will allow to essentially adjust frequency and external Q independently.

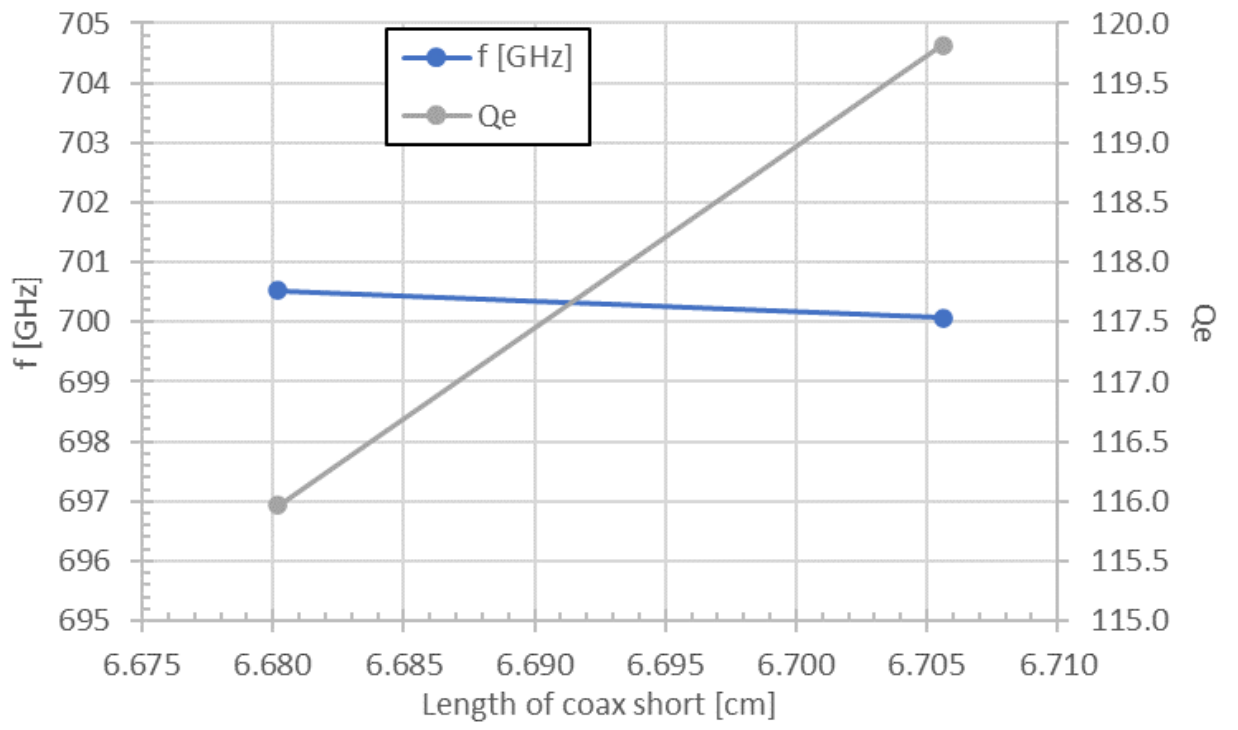

Fig. 25: Simulated resonant frequency and external Q of the output cavity versus length of the coaxial short.

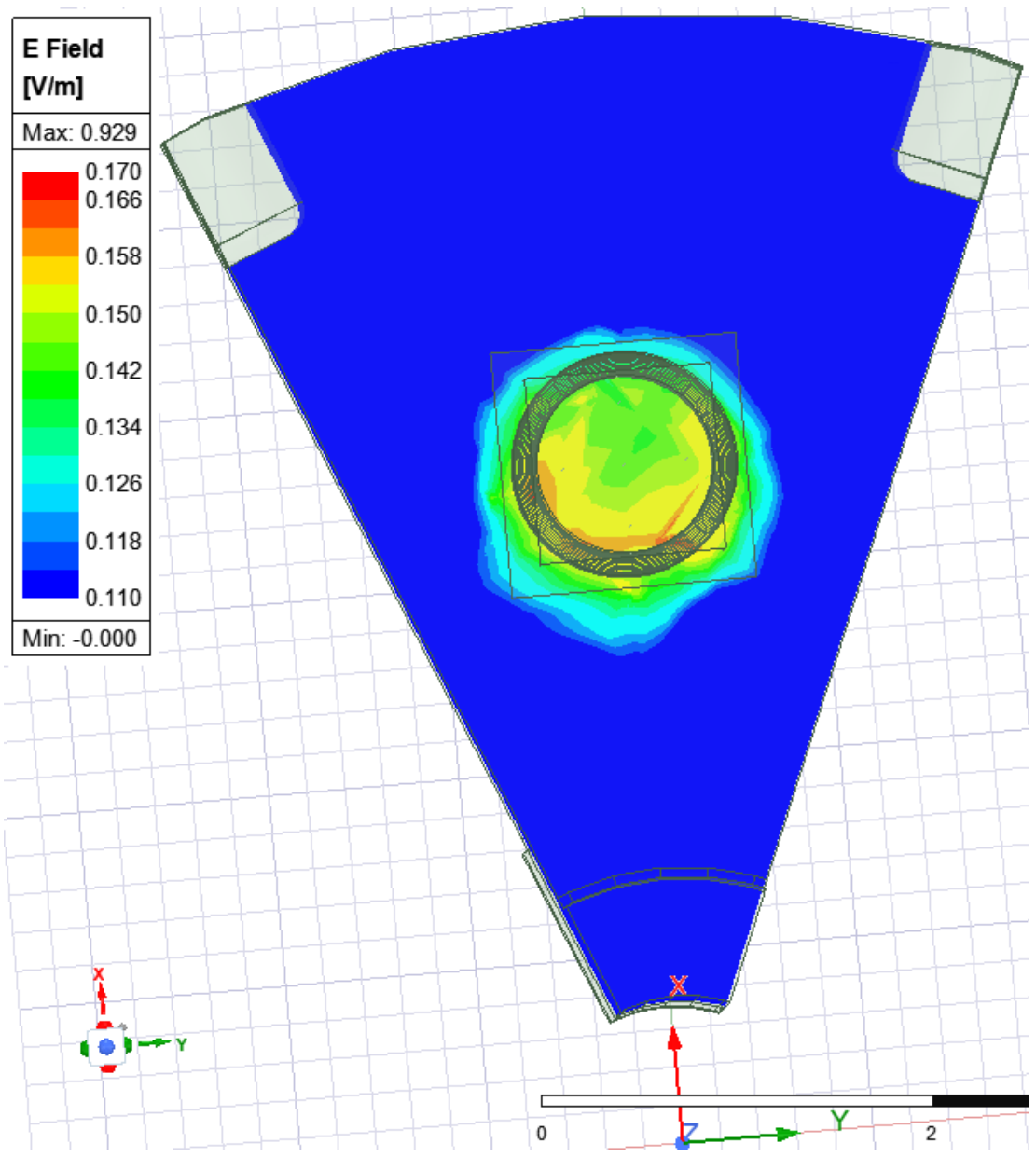

Fig. 26: Contour plot of the normalized electric field (1/8$^{th}$ slice) on a plane through the middle of the interaction gap.

Using a coaxial line in the center of the output cavity ensured that the RF electric fields at all eight beam locations are the same, thus allowing uniform extraction of the RF power from all the beams. There is however some non-uniformity within the area of each beam since the electric field is higher towards the "inside" edge of the beam (beam edge closer to the center of the cavity) compared the "outside" edge of the beam. This is shown in the electric field contour plot in Fig. 26. Unfortunately, there are some limitations on how close to the cavity center the beams can be placed or how large the inner conductor can be made.

HFSS simulations with driven modal solution types were performed to ensure that the RF electric field doesn't exceed acceptable limits. The results in Fig.27 show a maximum electric field of 8.7 MV/m.

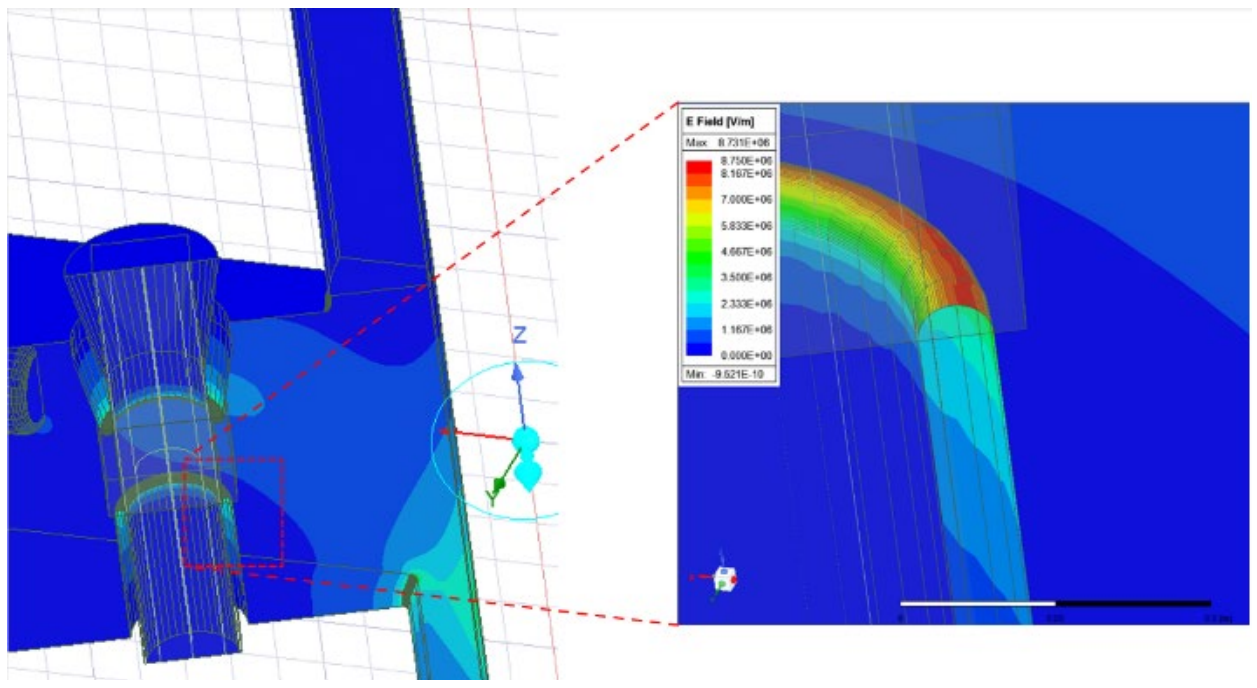

Fig. 27: HFSS model showing a 1/16th slice of the output cavity. The results depict the electric field values on the surfaces (and the symmetry plane in the background). The enlarged inset shows the area with the highest voltage gradients on the radius of the upstream drift tube "nose".

All major components and subassemblies of the MBIOT have been modeled in Solidworks. Figure 28 shows the model of the output cavity and the bottom part of the collector.

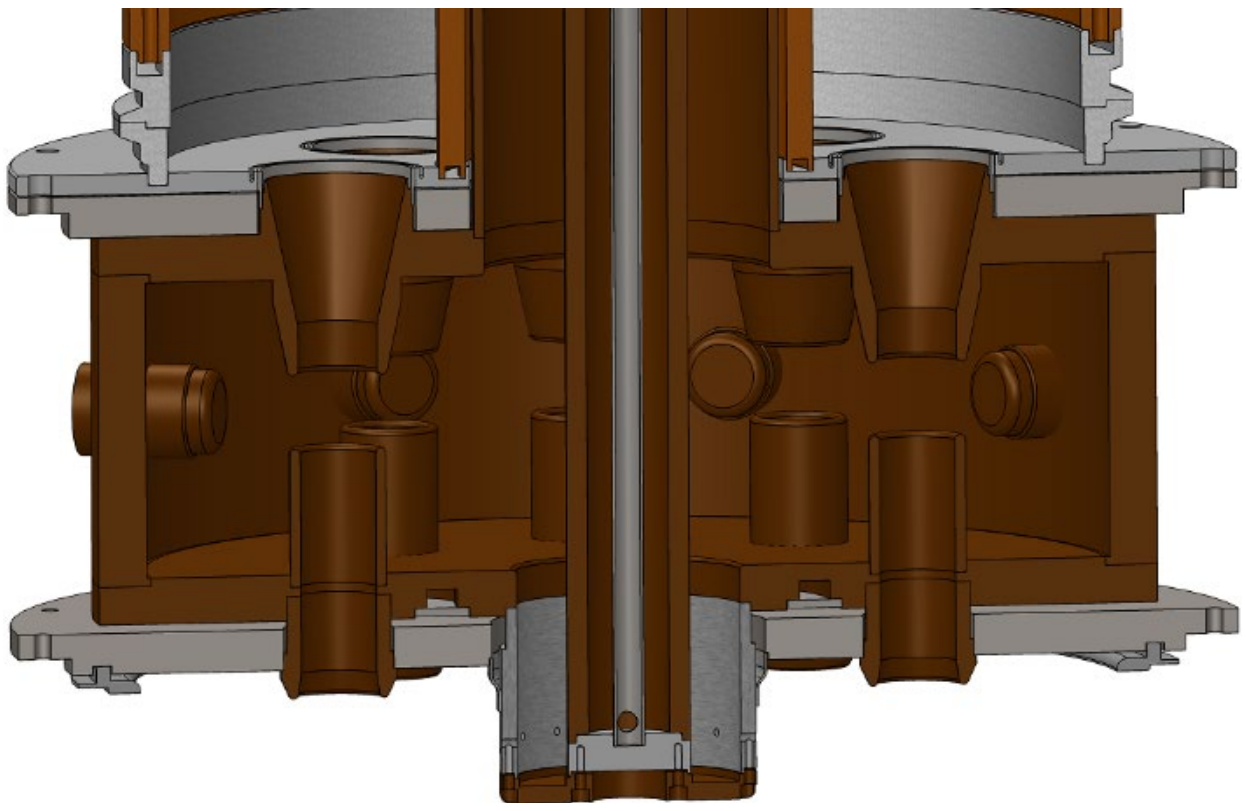

Fig. 28: Cross-sectional view (Solidworks model) of the MBIOT output cavity.

The RF exits the MBIOT output cavity through a coaxial line and a coaxial output window. The RF design of the output window and the transition to a standard 4-1/8" coax was created using the code CASCADE. Figure 29 shows the output window geometry in a cross-sectional view. The chart in Fig, 30 shows a low voltage standing wave ratio (VSWR) for this window design at the 700 MHz operating frequency.

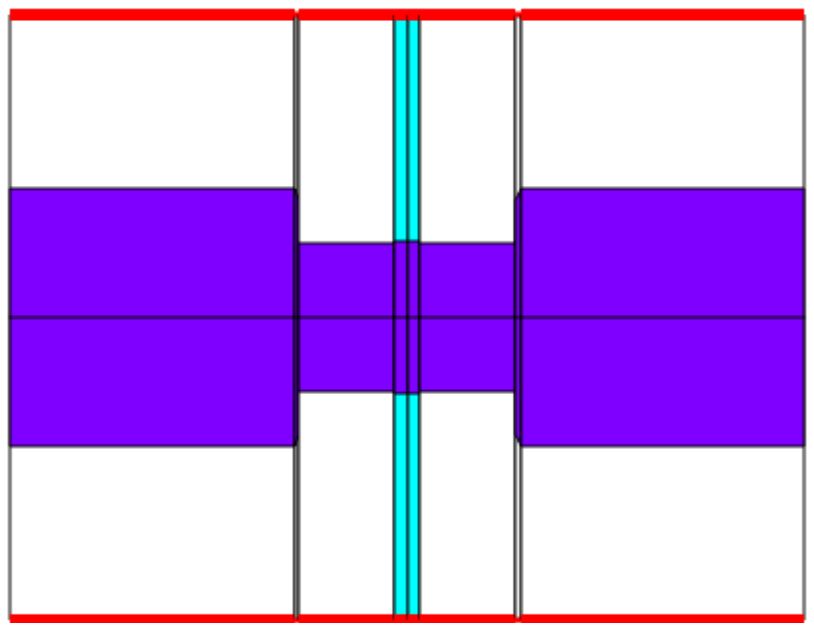

Fig. 29: RF Output Window geometry.

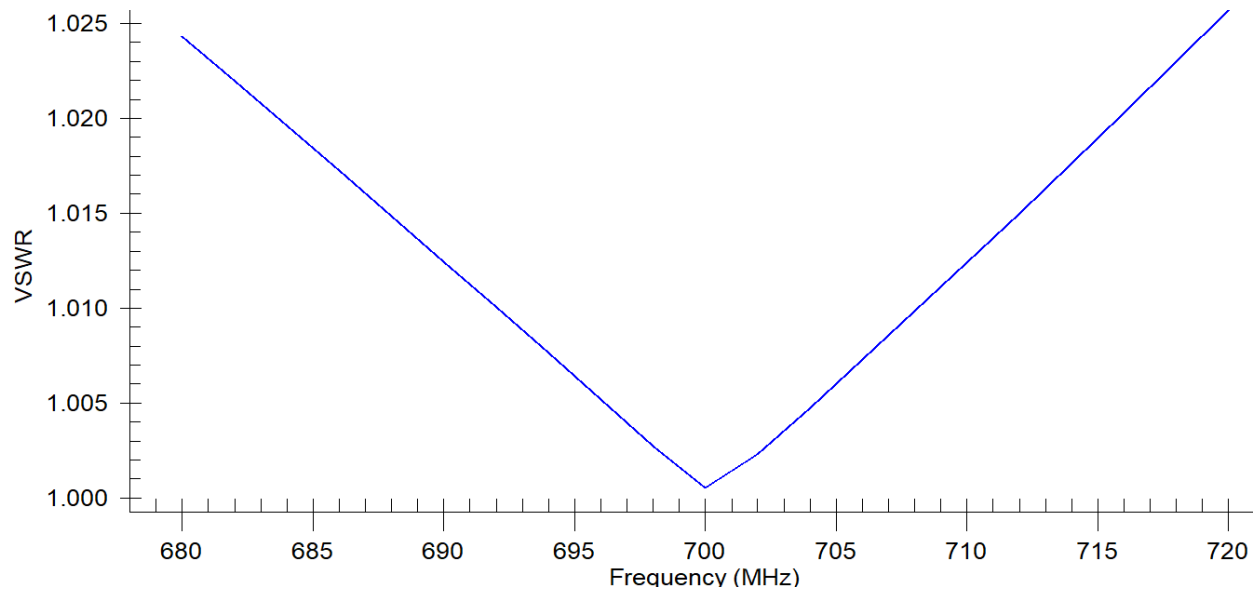

Fig 30: RF Output Window, VSWR versus frequency simulated using CASCADE.

## IV. SUMMARY AND CONCLUSION

In this paper, we have described whole-cavity simulations with NEMESIS and component design for an 8-beam MBIOT with the inclusion of a 3rd harmonic component on the drive. The numerical formulation is an outgrowth of the NEMESIS simulation code [1.9] which has been extended by the incorporation of a three-dimensional Poisson solver based upon the Petsc package available from Argonne National Laboratory.

It was found that the effect of the 3rd harmonic on the efficiency is greatest when the phase of the 3rd harmonic is shifted by $\pi$ radians with respect to the fundamental drive signal and with 3rd harmonic powers greater than about 50% that of the fundamental drive power [9]. For the present example, we show that efficiencies approaching 82% are possible by this means and that improvements in the performance due to the 3rd harmonic can reach 8.5%.

We remark that the NEMESIS formulation with the incorporation of the three-dimensional Poisson solver allows for the whole-cavity simulation of multi-beam klystrons as well. Application of NEMESIS to the simulation of a multi-beam klystron under development at Calabazas Creek Research will be reported in a future paper.

We also present the design and analysis of a compact input coupler suitable for both fundamental and 3rd harmonic drive of the electron gun grid. The input coupler is tunable from 698-703 MHz.

The program also investigated a moly grid to replace the existing PG grid used in IOT guns. The analysis showed that the moly grid would operate at a lower temperature, though compensation is required for the difference in thermal expansion.

Finally, we designed an output cavity to extract energy from the eight electron beams and produce RF power extracted through a coaxial window on the tube axis. The design incorporates tuning plugs to ensure precise frequency control in the assembled output cavity.

This device, if successfully built and tested, would represent a significant advanced in RF source technology in this frequency range. It would provide a higher efficiency, more compact, and lower cost alternative to existing RF sources.

## ACKNOWLEDGEMENTS

This material is based upon work supported by the U.S. Department of Energy, Office of Science, Office of High Energy Physics, under Grant No. DE-SC0019800.

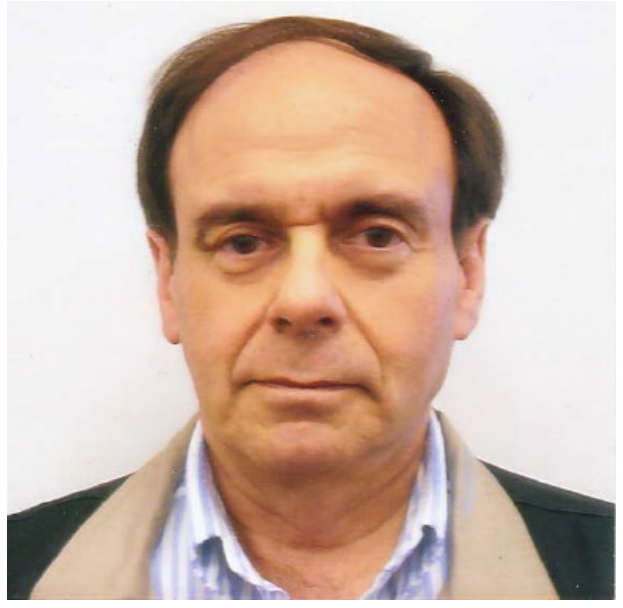

**Dr. H.P. Freund** is a theoretical plasma physicist currently studying coherent radiation sources such as free-electron lasers, traveling wave tubes, klystrons and inductive output tubes, and cyclotron, Cerenkov, and orbitron masers by both analytical and numerical methods. He has published more than 180 papers in refereed journals, made numerous contributions to books and published proceedings, and coauthored a book entitled *Principles of Free-electron Lasers* [Springer, Cham, Switzerland, 2018, 3rd edition]. In addition to this scholarly activity, Dr. Freund has also made contributions to more popular scientific literature with contributions on free-electron lasers published in *Scientific American* magazine and the Academic Press Encyclopedia of Science and Technology. The article in *Scientific American* has been translated and republished in *Veda a Technika* (Science and Technology), which was published by the Czechoslovak Academy of Sciences. Dr. Freund has extensive experience in the simulation of microwave tubes. He has developed important simulation codes for treating free-electron lasers and masers, cyclotron masers and gyrotrons, traveling wave tubes, and klystrons and inductive output tubes.

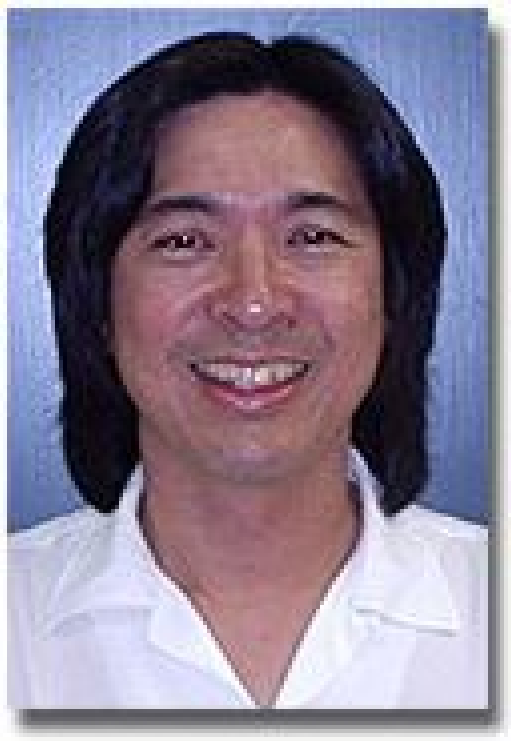

**Thuc Bui** received his M.S. from the University of California at Berkeley and Engineer degree in Applied Mechanics from Stanford University. He performed extensive research in finite element methods, developed and implemented three-dimensional, linear and quadratic tetrahedral and hexahedral elements to solve problems in elasticity, compressible fluids, supersonic viscous flows, electrostatic, magnetostatic and Helmholtz fields. He uses object-oriented programming techniques, design patterns and finite element methods to solve charged particle beam problems. He is the principal author of Beam Optics Analyzer (BOA) and was responsible for the development of the particle pusher, emission algorithms, Poisson and Curl-Curl field solvers. The particle pusher and the field solvers, using the finite element method, are intimately integrated with the framework originally developed by RPI, Simmetrix and further improved by Mr. Bui to utilize its novel and world class adaptive meshing facility. For the Curl-Curl solver in magnetostatic and Helmholtz problems, he implemented the vector finite element method to strongly enforce the tangential continuity of the field. In collaboration with NCSU, Mr. Bui implemented the Monte Carlo method with a secondary emission model of elastically backscattered electrons from the first principle. He then extended this model to simulate the power density field inside the solid due to backscattered electrons for heat transfer analysis. He also used iterative methods Nelder-Mead and Implicit Filtering in BOA to optimize the emitter shape, size, electrode spacing and focusing magnetic field to achieve to desired current and beam shape. He has recently completed a parallelized 3D Poisson solver using the finite difference method in Nemesis to push particles in a multiple beam devices.

**Dr. R. Lawrence Ives** (M’82-SM’90-LM’16) received his Ph.D. from N.C. State University in plasma physics in 1984 and began his microwave career in the Gyrotron Department at Varian Associates, Inc. in Palo Alto, CA. In that position he was responsible for designing electron guns, gyrotron circuits, collectors, and waveguide components.

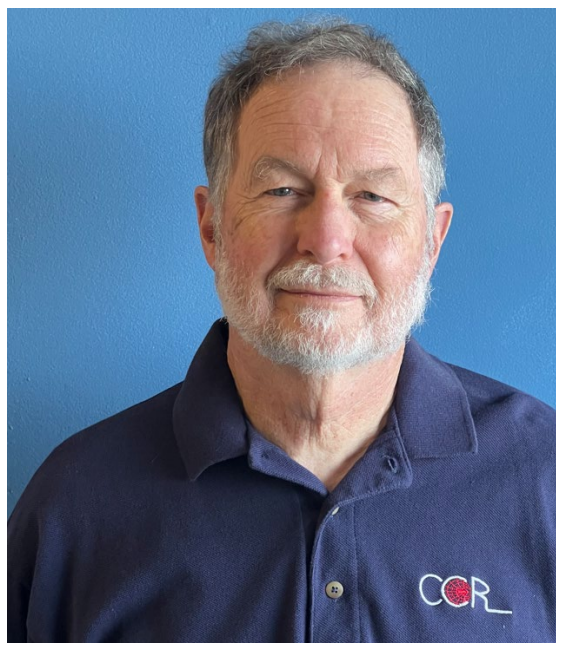

He founded Calabazas Creek Research, Inc. in 1994, which is involved in software development, microwave tube and component design, and uses of microwave power for environmental, research, defense, and heating applications. Dr. Ives was principal investigator on programs to develop long life, high current density cathodes, long life, high quantum efficiency photocathodes, high power RF windows, and multiple beam klystrons. Current areas of research include use of Atomic Layer Deposition to reduce cooling channel corrosion, direct coupled gyrotrons, multiple beam triode amplifiers, and application of additive manufacturing to RF sources and components.

Dr. Ives also provides consulting support to several commercial companies on electron guns, cathodes, and RF source development.

Dr. Habermann received a Ph.D. (Dr. rer. nat.) in Physics

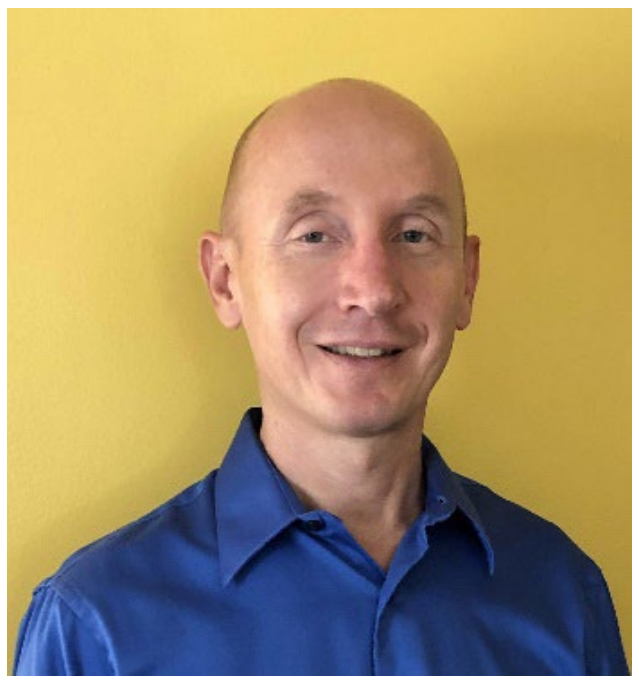

from the University of Wuppertal (Germany) in 1999, where he conducted research in field emission. He investigated carbon nanotubes, CVD diamond thin films, diamond-coated silicon microstructures for cold cathode applications, as well as researched the basic mechanisms of 'parasitic' field emission impacting superconducting RF cavities used in particle accelerators. From 2000 to 2019 he worked at Communications and Power Industries in Palo Alto, CA, on the development and manufacturing of klystrons. Dr. Habermann managed multiple programs to develop, produce and deliver prototypes or initial production runs of new vacuum electron devices. He contributed to numerous programs within the company and provided engineering support in production. In 2019 Dr. Habermann joined Calabazas Creek Research where he is supporting various R&D programs. He is also responsible for the company's Quality Assurance.

Dr. Read received his Ph.D. in 1975 from Cornell University in Electrical Engineering and Plasma Physics. His graduate work was on the generation and propagation of Intense Relativistic Electron Beams (IREBs). Following his postdoctoral work, Dr. Read worked at the Naval

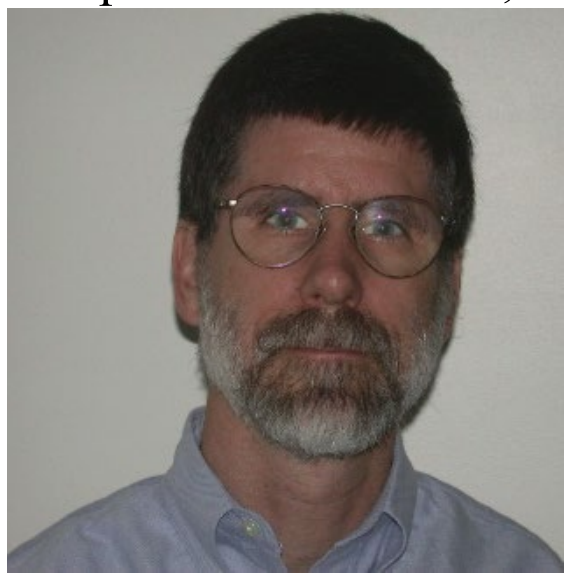

Research Laboratory (NRL) and from 1983 to 1986 was head of the Gyrotron Oscillators and Plasma Interactions Section of Plasma Physics Division. In 1986, he became a Principal Research Engineer with Physical Sciences Inc., where he was responsible for research programs on microwave and electron beam generation and their applications. In 1999, Dr. Read joined Calabazas Creek Research, where he is responsible for electron gun and high power microwave device research. He is currently Chief Scientist at CCR.